\documentstyle[12pt,epsf,epsfig]{article}
\def\be{\begin{equation}}
\def\ee{\end{equation}}
\def\ba{\begin{eqnarray}}
\def\ea{\end{eqnarray}}

\def\bdm{\begin{displaymath}}
\def\edm{\end{displaymath}}

\def\ga{~\mbox{\raisebox{-.6ex}{$\stackrel{>}{\sim}$}}~}
\def\bq{\begin{quote}}
\def\eq{\end{quote}}

 at 10truept

\newcommand{\beq}{\begin{equation}}
\newcommand{\eeq}{\end{equation}}
\newcommand{\bea}{\begin{eqnarray}}
\newcommand{\eea}{\end{eqnarray}}
\newcommand{\beqa}{\begin{eqnarray}}
\newcommand{\eeqa}{\end{eqnarray}}

\def\ga{~\mbox{\raisebox{-.6ex}{$\stackrel{>}{\sim}$}}~}

\def\ltap{\ \raise.3ex\hbox{$<$\kern-.75em\lower1ex\hbox{$\sim$}}\ }
\def\gtap{\ \raise.3ex\hbox{$>$\kern-.75em\lower1ex\hbox{$\sim$}}\ }
\def\gl{\ \raise.5ex\hbox{$>$}\kern-.8em\lower.5ex\hbox{$<$}\ }
\def\roughly#1{\raise.3ex\hbox{$#1$\kern-.75em\lower1ex\hbox{$\sim$}}}

\begin{document}

\thispagestyle{empty}
\begin{flushright}
arXiv:yymm.nnnn [hep-th]\\
June 2011
\end{flushright}
\vspace*{.65cm}
\begin{center}
{\Large \bf Galileon Hairs of Dyson Spheres, Vainshtein's Coiffure}
\vskip.3cm
{\Large \bf and Hirsute Bubbles}\\

\vspace*{1.2cm} {\large Nemanja Kaloper$^{a,}$\footnote{\tt
kaloper@physics.ucdavis.edu}, Antonio Padilla$^{b,}$\footnote{\tt
antonio.padilla@nottingham.ac.uk} and Norihiro Tanahashi$^{a,}$\footnote{\tt
tanahashi@ms.physics.ucdavis.edu}}\\
\vspace{.5cm} {\em $^a$Department of Physics, University of
California, Davis, CA 95616, USA}\\
\vspace{.2cm} {\em $^{b}$School of Physics and Astronomy, 
University of Nottingham, Nottingham NG72RD, UK}\\

\vspace{1.cm} ABSTRACT
\end{center}
We study the fields of spherically symmetric thin shell sources, a.k.a. Dyson spheres, in a {\it fully nonlinear covariant} theory of gravity with the simplest galileon field. We integrate exactly all the field equations once, reducing them to first order nonlinear equations. For the simplest galileon, static solutions come on {\it six} distinct branches. On one, a Dyson sphere surrounds itself with a galileon hair, which far away looks like a hair of any Brans-Dicke field. The hair changes below the Vainshtein scale, where the extra galileon terms 
dominate the minimal gradients of the field. Their hair looks more like a fuzz, because the galileon terms 
are suppressed by the derivative of the volume determinant. It shuts off the `hair bunching' over the 
`angular' 2-sphere. Hence the fuzz remains dilute even close to the source. This is really why the 
Vainshtein's suppression of the modifications of gravity works  close to the source.
On the other five branches, the static solutions are all {\it singular} far from the source, and shuttered off from asymptotic infinity. One of them, however, is really the self-accelerating branch, and the singularity is removed by  turning on time dependence. We give examples of regulated solutions, where the Dyson sphere explodes outward,  and its self-accelerating side is nonsingular. These constructions may open channels for nonperturbative transitions between branches, which need to be addressed further to determine phenomenological viability of multi-branch gravities. 

\vfill \setcounter{page}{0} \setcounter{footnote}{0}
\newpage

\tableofcontents

\vskip1cm

\section{Introduction}

Recently there have been numerous attempts to deform General Relativity with the addition of extra operators which reside in the infrared (for a review, see \cite{tonyreview}). 
The main purpose was to see if such deformations may yield novel avenues to address the cosmological constant problem. In the process, it was realized that such deformations generically bring in new degrees of freedom in the gravitational sector, by breaking residual gauge symmetries of General Relativity. 
The new theories, which are not trivial
rearrangements of a scalar-tensor gravity in some cumbersome gauge, display a phenomenon of strong coupling, which at the classical level leads to  the breakdown of the perturbative description of the extra modes at very large distances, on backgrounds with heavy sources.
A prototypical example of such behavior was already encountered in Fierz-Pauli massive gravity \cite{pauli},  where  the extra mode(s) do not decouple even when the graviton mass is taken  to zero \cite{discontinuity}, and where  nonlinear effects  become big at  distances that can be much larger than the Schwarzschild radius of the source \cite{vainshtein}. The latter observation,
made by Arkady Vainshtein \cite{vainshtein}, prompted him to suggest that the nonlinear effects might conspire and tame the extra modes in the graviton spectrum at distances shorter than the scale where linearized perturbation theory breaks down. Similar behavior has been argued \cite{dvali} to also arise in other models of modified gravity such as DGP \cite{dgp}, 
galileon theory \cite{galileon}, and the new model of massive gravity recently proposed by de Rham and Gabadadze \cite{derham} (see, for example, \cite{gustavo, nyu, gruzinov2}).

However a complete demonstration of how and why this effect works is lacking.
 There were attempts to integrate background equations in purported nonlinear extensions of Fierz-Pauli theory \cite{damour,deffayet},  but they have conflicting outcomes.  In \cite{damour}, it was noted that the solutions which are flat at infinity do not connect onto the smooth interiors near sources, but become singular in between the desired asymptotic regions. More recently, in \cite{deffayet}, numerical exploration suggested otherwise, and those authors claimed that the solutions which extrapolate smoothly between mass sources and asymptotic infinity do exist. Nevertheless, since this work was based on what is really not a complete and fully covariant theory, one may worry if the additional terms may have been omitted which could change the outcome of the numerics. Indeed, even those authors warn of the sensitivity of the perturbative approaches on nonlinear effects. Moreover, even if that is not so, a more explicit and direct understanding of this phenomenon is warranted, as much of the work on nontrivial deformations of General Relativity at some level utilizes the so-called ``Vainshtein mechanism". One would therefore wish to have a foolproof demonstration of operability of the said mechanism, in a way that leaves little doubt as to what may be going on.

That is the purpose of this work. Here we shall explore the simplest theory which features the Vainshtein mechanism, the fully nonlinear covariant cubic galileon, with conformal couplings to matter \cite{covgal}.
To simplify the analysis, we will focus on the fields of spherically symmetric thin shell sources, or Dyson spheres \cite{dyson}. Thus the effects of matter sources on the fields will be modeled by the boundary conditions which the fields satisfy across matter shells. This simplification does not reduce the generality of our results, but it does enable us  to integrate exactly all the field equations once in the static case. 
Hence the resulting field equations governing static solutions reduce to first order nonlinear equations. For the cubic galileon, static solutions come on {\it six} distinct branches. 

On one, a Dyson sphere surrounds itself with  galileon hair, which far away looks like a hair of any Brans-Dicke field, with  its usual long strands, $\phi \sim 1/r$, governed by the standard $4D$ Gauss law.
The hair changes  at  the Vainshtein scale, from wherein the extra galileon terms dominate the
minimal gradients of the field. In contrast to the long canonical hair, their hair looks more like a fuzz, because the galileon terms are suppressed by the derivative of the volume determinant. This shuts off the `hair bunching' over the `angular'  2-sphere at shorter distances. Hence the short fuzz is less dense even close to the source. We stress that this is different
from screening in gauge theories, where the medium develops spontaneous polarization which shields the fields generated by sources. Instead, it's the nature of the hair which changes, while the Gauss law formula and the force felt by probes remain the same. The transition between the two hair regimes is controlled by the dimensional coupling constant of the cubic galileon term, which sets up the Vainshtein scale 
$r_V$. When the mass scale of this coupling is chosen to be low, the distance $r_V$ is long, implying that the hair fraying occurs far from the source, where the hair is dilute and hence the extra force field from the scalar is weak. The diluted fuzz ensures that the field strength can not grow very large further inward.  This is really why the Vainshtein's suppression of the modifications of gravity works close to the source, as we will discuss in detail. In turn, the price one pays is that the effective Lagrangian for the galileon is cut off at a low scale, albeit we confirm that in terrestrial conditions this cutoff is close to a millimeter if one picks the galileon couplings such that they could yield interesting cosmological effects at very large distances.

On the other five branches, the smooth static Vainshtein's coiffure does not stick on.  All the static solutions on these additional branches are {\it singular} far from the source, and closed off from asymptotic infinity. We show this qualitatively, but the proof is general, as we use the phase space analysis of the system, where the singularities manifest as the bifurcations of the phase space trajectories. Four branches of singular solutions are descendants from similar singular solutions in Brans-Dicke gravity \cite{old,bmq}. Yet, the fifth is really the self-accelerating galileon sector where the singularity is a manifestation of forcing the geometry to stay static. Static solutions in the self-accelerating sector are inconsistent, since the scalar condensate far away drives time evolution. This is reminiscent to what happens, for example, in the case of singular global strings
\cite{cohkaplan}, which become regular when allowed to inflate in the core \cite{ruth}.  The difference is that while the scalar hair for global strings is fully supported by the string core, here the local sources merely trigger the scalar gradient. The faraway evolution is controlled by the playoff of the scalar terms, with only minor contributions from the sources. 
Although we do not analyze the Vainshtein phenomenon on the regulated time dependent self-accelerating branch, we expect that in the simple model which we analyze cannot really yield a healthy dynamics there, because the scalar field on that branch is a ghost. So while the short range hairs of this ghost may be suppressed, its long range behavior would still yield pathologies. On the other hand, if the ghost may be excised by further modifying the theory, and leaving the self-acceleration on smooth backgrounds, we believe that the Vainshtein effect may remain operable on such branches too. Yet, it remains to be seen if multi-branch theories without any ghosts anywhere actually exist.

Motivated by the similarity with singular global strings, we construct explicit examples, in thin wall limit, of regular time-dependent solutions in our covariant cubic galileon theory.
 In such configurations, the Dyson sphere is really an expanding bubble wall, separating self-accelerating branch from a (section of) asymptotically flat branch, and vice versa.
Such configurations can be viewed as bubbles of one vacuum formed inside another, and are channels for nonperturbative transitions between branches. Our bubbles indicate that the self-accelerating branch is unstable to decay to the asymptotically flat branch, with the transitions described by nucleation of bubbles with positive tension. The bubbles describing pockets of self-accelerating branch inside the Minkowski vacuum, however, require negative tension, and it remains to see if such bubbles can be consistently resolved within the energy spectrum of the theory itself \cite{tony&co,koyama-tanaka-pujolas}. This is particularly interesting because
such non-perturbative channels may occur in other modified gravities with strongly coupled modes and multiple classical branches of solutions \cite{tony&co}, particularly if they reduce to galileon models in the decoupling limit. Hence it is important to study this issue further, to determine phenomenological viability of such models. The galileon models offer a setup within which this important and interesting question may be addressed efficiently.

\section{Setup}

 As we already stressed in the introduction, we will work with the cubic galileon theory with  conformal couplings to matter,
 and restrict matter sources to reside on the surface of Dyson spheres. 
 This is described by the following action \cite{covgal},
\be
S = \int d^4 x \left\{\sqrt{-g} \left(\frac{M_{Pl}^2}{2} R - \frac12 (\partial \phi)^2 - \hat \alpha (\partial \phi)^2 \nabla^2 \phi \right) 
 - \sqrt{-\hat g} {\cal L}_{\rm matter}(\hat g^{\mu\nu}, {\rm matter ~fields} )  \right\} \, , \label{action}
\ee
where $\hat g_{\mu\nu} = e^{\beta \phi/M_{Pl}} g_{\mu\nu}$. 
We choose the overall sign of the leading order galileon term, $- \frac12 (\partial \phi)^2$, to ensure a stable ghost-free Lorentz-invariant vacuum. It then follows that the so-called self-accelerating vacuum is plagued by ghosts.  In addition, we assume, without loss of generality, that $\hat \alpha > 0$\footnote{If $\hat \alpha < 0$ we let $\phi \to -\phi$ and $\beta \to -\beta$ to recover a theory with $\hat \alpha > 0$.}, but allow $\beta$ to take either sign.
This action can
now  be split into the ``bulk sector", off the Dyson spheres, and the 
sector residing on the spheres, where the equations simply reproduce Israel conditions relating the geometries inside and outside the Dyson sphere. Away from them, the simplest approach to setting up the bulk problem for the static background is to work in the action on spherically symmetric backgrounds. When the background geometry is taken to be static and spherically symmetric,
\ba
ds^2 &=& - e^{2\nu} dt^2 + e^{2\lambda} dr^2 + e^{2\sigma} d\Omega_2 \, , \nonumber \\
\phi &=& \phi(r) \, , 
\label{metricmatter}
\ea
we can dimensionally reduce the bulk action for the gravitational sector with the galileon on the Killing vectors
$\partial_t, \partial_\theta, \partial_\phi$. A straightforward calculation yields the following $1D$ effective 
action
\be
S^{bulk}_{\rm eff} = \int dr \left\{ (2\sigma' \nu' + \sigma'^2 ) e^{\nu+2\sigma - \lambda} 
+ e^{\nu+\lambda}
- \frac12 e^{\nu+2\sigma-\lambda} \varphi'^2 
- \frac{2 \alpha}{3} (\nu' + 2\sigma') \varphi'^3 e^{\nu+2\sigma-3\lambda} \right\} \, .
\label{1daction}
\ee
The primes denote $r$-derivatives. We have divided the whole action by $M_{Pl}^2$, redefined the
scalar by $\phi = M_{Pl} \varphi$, the galileon coupling by $ \alpha = \hat \alpha M_{Pl}$, and dropped the 
irrelevant overall factor of angular area $4\pi$. In arriving at Eq. (\ref{1daction}) we have used the fact that the
cubic galileon term, $\sqrt{-g} (\partial \varphi)^2 \nabla^2 \varphi$ is really only a product of first derivatives, up to an irrelevant boundary term, because on the background (\ref{metricmatter}) it becomes $\frac13 e^{-2\nu-4\sigma} (\varphi'^3)'$. 

Varying this action with respect to the metric functions $\nu,\lambda,\sigma$ and the scalar $\varphi$ we find the equations of motion in the bulk. Indeed, the full system is
\ba
&& (e^{\nu+2\sigma-\lambda} \varphi')' + 2\alpha \left((\nu'+2\sigma') \varphi'^2 e^{\nu+2\sigma-3\lambda} \right)' = 0 \, ,\nonumber \\
&& \varphi'^2 + 4\alpha (\nu' + 2\sigma') \varphi'^3 e^{-2\lambda} = 2(2 \nu ' \sigma' + \sigma'^2) - 2 e^{2\lambda -2\sigma} \, ,\nonumber \\
&& (\sigma' e^{\nu+2\sigma-\lambda})' - \frac{\alpha}{3} (\varphi'^3 e^{\nu+2\sigma-3\lambda})' = 
 \frac12 (2\nu' \sigma' + \sigma'^2) e^{\nu+2\sigma-\lambda} + \frac12 e^{\nu+\lambda} - \frac14 e^{\nu+2\sigma-\lambda} \varphi'^2 \nonumber \\
&&~~~~~~~~~~~~~~~~~~~~~~~~~~~~~~~~~~~~~~~~~~~~~~~~~~~~~~~~~~~~~~~~~~~~ 
- \frac{\alpha}{3} (\nu'+2\sigma') \varphi'^3 e^{\nu+2\sigma - 3\lambda} \, ,
 \nonumber \\
&& [(\nu'+ \sigma') e^{\nu+2\sigma-\lambda}]' - \frac{2\alpha}{3} (\varphi'^3 e^{\nu+2\sigma-3\lambda})' = 
(2\nu' \sigma' + \sigma'^2) e^{\nu+2\sigma-\lambda} - \frac12 e^{\nu+2\sigma-\lambda} \varphi'^2 \nonumber \\
&&~~~~~~~~~~~~~~~~~~~~~~~~~~~~~~~~~~~~~~~~~~~~~~~~~~~~~~~~~~~~~~~~~~ 
- \frac{2\alpha}{3} (\nu'+2\sigma') \varphi'^3 e^{\nu+2\sigma - 3\lambda} \, ,
\label{systemfull}
\ea
where the equations are results of varying (\ref{1daction}) by $\varphi, \lambda, \nu, \sigma$, respectively.
It is obvious that the former two equations are first order (after trivially integrating the first equation). The latter
two look more formidable. However, subtracting twice the third equation from the fourth, we find
\be
[(\nu'-\sigma') e^{\nu+2\sigma-\lambda}]' = - e^{\nu+\lambda} \, . \label{difference}
\ee
Now, since we are at liberty to gauge fix the ``lapse function" $\lambda$ at will by picking an appropriate 
coordinate transformation $ r \rightarrow f(r)$, we can choose it such that $\lambda = - \nu$. In this gauge Eq. (\ref{difference}) reduces to $[(\nu'-\sigma') e^{\nu+2\sigma-\lambda}]' = - 1$, and so it is also readily integrable. It suffices 
to keep only it and the first two of the system (\ref{systemfull}) because by the covariance of the theory, the four equations
in (\ref{systemfull}) are not all independent. One of them is a linear combination of the (derivatives of the) others, 
as enforced locally by Bianchi identities and stress-energy conservation. 

Therefore in the $\lambda = - \nu$ gauge the full system in the bulk reduces to a first order system of three differential equations,
\ba
&& e^{2\nu+2\sigma} \varphi' + 2\alpha(\nu'+2\sigma') \varphi'^2 e^{4\nu+2\sigma} = Q \, ,\nonumber \\
&& \varphi'^2 + 4\alpha (\nu' + 2\sigma') \varphi'^3 e^{2\nu} = 2(2 \nu ' \sigma' + \sigma'^2) - 
2 e^{-2\nu -2\sigma} \, ,\nonumber \\
&& \nu' = \sigma' -  e^{-2\nu - 2\sigma} (r-q) \, ,\label{system}
\ea
where $Q$ and $q$ are two integration constants, to be determined by the boundary conditions to follow. The fourth equation
from (\ref{systemfull}) is redundant as noted, being a linear combination of these, and their first derivatives.

On to the matter sources on Dyson spheres. As we noted, they just induce sources 
in the Israel junction conditions relating the bulk configurations inside and out. Thus we can incorporate 
them in the standard way. However, in the spherically symmetric static case, we can use the shortcut, 
by employing the Einstein-galileon field equations with a $\delta$-function matter sources and using the Gaussian pillbox
trick to relate the singular terms in the bulk equations and compute the jumps in field first derivatives across the shells, while 
demanding continuity of the metric and scalar fields. The reason is that in the spherically symmetric 
static systems, the metric and scalar (\ref{metricmatter}) are already written in almost the Riemann normal form. 
Further, we only need to obtain three independent boundary conditions because the second equation in (\ref{system}) was already of the first order in the bulk form (\ref{systemfull}), and so it does not yield nontrivial junction conditions. 

To proceed, we note that the shell stress-energy in our gauge is given by
\be
T^\mu{}_\nu ({\rm matter}) =  {\rm diag}(-\rho, 0 , p, p) \delta(r-  r_0) \, , \label{shellstress}
\ee
where $\rho$ and $p$ are the energy density and the tangential pressure at the shell, and $r_0$ the shell radius. 
We have normalized the energy scales in the matter sector on the shell
relative to the local time coordinate $\hat t = e^{\nu_0} t$, because this absorbs away the local redshift of the matter sector 
coming from the gravitational potential well induced by the shell's own mass. In other words, $\rho$ and $p$ are {\it proper} energy density and pressure. The radial pressure 
vanishes because, as we noted, the radial Einstein-galileon equation is first order in derivatives, as 
in ordinary General Relativity. The two nontrivial conditions,  come from the ``$00$" and, say, the ``$22$" equations of the following,
\be
G^\mu{}_\nu = T^\mu{}_\nu(\varphi) + T^\mu{}_\nu({\rm matter}) \, ,
\label{einsteingal}
\ee
where we have normalized all dimensional scales to $M_{Pl}$. They are given by,
\ba
&& 2 \sigma'' e^{2\nu}  = -  \rho \, \delta(r-r_0) + \frac{2\alpha}{3} e^{4\nu} (\varphi'^3)' 
+ {\rm regular~terms} \, , \nonumber \\
&& (\nu'' + \sigma'') e^{2\nu} =   p \, \delta(r-r_0) + \frac{2\alpha}{3} e^{4\nu} (\varphi'^3)' 
+ {\rm regular~terms} \, . \label{einsteinjunc}
\ea
Meanwhile the scalar field equation,
\be
\nabla^2 \varphi + \alpha \left(2 \nabla_\mu(\nabla^2 \varphi \nabla^\mu \varphi) - \nabla^2 (\nabla \varphi)^2 \right) +  
\frac{\beta}{2} T^\mu{}_\mu({\rm matter}) = 0 \, ,
\label{scalar}
\ee
yields
\be
(e^{2\nu+2\sigma} \varphi')' + 2 \alpha \left((\nu'+2\sigma') \varphi'^2 e^{4\nu+2\sigma}\right)' + \frac{\beta}{2}
e^{2\sigma} (2p-\rho) \delta(r-r_0) = 0 \, . \label{scalarjunc}
\ee
Integrating (\ref{einsteinjunc}), (\ref{scalarjunc}) across the shell, between $r_0\pm \epsilon$ and taking $\epsilon \rightarrow 0$ 
finally gives us the required junction conditions. With all the functions evaluated at $r=r_0$, they are
\ba
&&\Delta \varphi' + 2\alpha e^{2\nu} \Delta \left( (\nu'+2\sigma') \varphi'^2 \right) 
=  \frac{\beta}{2} e^{-2\nu} (\rho - 2p) \, , \nonumber \\
&& \Delta \sigma' = -  e^{-2\nu} \frac{\rho}{2} + \frac{\alpha}{3} e^{2\nu} \Delta (\varphi'^3) \, , \nonumber \\
&& \Delta \nu' + \Delta \sigma' =  e^{-2\nu} p + \frac{2\alpha}{3} e^{2\nu} \Delta (\varphi'^3) \, .
\label{bcs}
\ea
Here the symbol $\Delta P$ designates the difference of the quantity $P$ between the different sides of the
Dyson sphere. We should stress that both the interior and exterior solutions must be appropriately gauge fixed to 
ensure that the Dyson sphere resides at the same value of the radial coordinate viewed from either side.

In any case, we see that to determine the gravitational and galileon fields surrounding a Dyson sphere we need
to find solutions of (\ref{system}) satisfying the junction conditions (\ref{bcs}) on the sphere. Let us turn to that now.

\section{Growing Hairs: Bulk Solutions} 

Note that the solution that describes the interior of a regular static stable Dyson sphere should be unique, a 
spherical section of flat Minkowski space with a constant galileon $\varphi$. The reason is that 
the static galileon configurations which continuously deform the Minkowski vacuum do not violate null energy condition. 
If we had turned them on in the interior, they would have induced a curvature singularity in the center, 
as dictated by Hawking-Penrose theorems. The singularity could be avoided by the null energy violating, time-dependent galileon configurations, but such backgrounds would be plagued by ghosts if the Minkowski vacuum is ghost-free, as in the example that we focus on. Therefore, the interior solutions degenerate to empty space. 

This should not be too surprising, since after all one encounters precisely the same situation in conventional General Relativity
in the presence of normal matter sources. Moreover, such configurations are also suggested by black hole no-hair theorems,
which require that black holes can't support primary scalar hairs, setting their charges to zero. Quite so, we note that the Schwarzschild solution is a solution of the cubic galileon theory if we set the integration constant $Q$ to zero, and pick the
trivial root of the first of Eq. (\ref{system}). Given this, we expect that such standard black holes should form by the collapse of 
Dyson spheres, if it is controllably slow and if during it scalar hair is shed away by radiation. It is hard to imagine that an interior which contained a lot of scalar hair could shed it quickly enough to let a completely regular black hole form. 
This reasoning is already highly suggestive that Vainshtein configurations should exist somewhere in the theory 
once we pick empty interiors.

Let us first consider the limiting case $\alpha = 0$ in (\ref{system}). In this case the theory reduces to the simple scalar-tensor Brans-Dicke theory, parameterized by the single coupling to matter $\beta$. The bulk equations (\ref{system}) are then exactly integrable, as is well known \cite{old,bmq}. The
solutions in our gauge $\lambda = - \nu$ can be written as \cite{bmq}
\ba
e^{2\nu} &=& \left(1-\frac{\ell}{r}\right)^\delta \, , \nonumber \\
e^{2\sigma} &=& r^2 \left(1-\frac{\ell}{r}\right)^{1-\delta} \, , \nonumber \\
e^{\varphi} &=& \left(1-\frac{\ell}{r}\right)^\gamma \, ,
\label{scalarsolns}
\ea
where $\ell, \delta, \gamma$ are integration constants, the latter two subject to the condition $\delta^2 + \gamma^2 = 1$.
Note that for $\gamma \ne 0$, the solution (\ref{scalarsolns}) in actual fact is a compact way of writing {\it two} separate families of solutions:
\begin{itemize} 
\item one family extrapolates between the curvature singularity at $r = \ell$ (which is really the ``center" of the coordinate system, as seen from the second of Eq. (\ref{scalarsolns}), because  there the area of the constant $r$ spheres vanishes), and Minkowski geometry at infinity;
\item another family extrapolates between two curvature singularities, 
$r=0$ and $r=\ell$. 
\end{itemize}
In this case the theory has the symmetry $\varphi \leftrightarrow - \varphi$, reflected in the solutions as the 
invariance under simultaneous transformation $\gamma \rightarrow - \gamma$, the radial shift $r \rightarrow r+\ell$, and the sign flip $\ell \rightarrow -\ell$. This symmetry is trivial on the first family of solutions at infinity, where the 
scalar $\varphi$ vanishes. The solutions which are regular at infinity flow to the fixed point of the discrete
symmetry $\varphi \leftrightarrow - \varphi$, given by $\varphi =0$.
To tailor together the full geometry, we need to pick the family of solutions based on asymptotic properties far from the source, and use the junction conditions on the shell to uniquely specify the specific solutions within this family. The latter, doubly-singular, family of solutions can be thrown away in conventional gravity, be it General Relativity or Brans-Dicke, because one of the two singularities would be far from the source, and the solution would not flow to the vacuum outside of matter. 
Junction conditions, specified with usual matter that satisfies conventional energy conditions, never forbid us from doing this. Furthermore, quantum 
effects do not reintroduce the singular solutions, because the transitions from regular backgrounds are suppressed by large Euclidean actions demanded by the singular regions. However, in the case of modified theories, in principle the junction conditions may alter these conclusions, and one may have to pay special attention to this issue.

Let us now return to (\ref{system}) with $\alpha > 0$. To classify the families of solutions which the system admits, we note that the discrete symmetry $\varphi \leftrightarrow - \varphi$ is explicitly broken by the simultaneous presence of
quadratic and cubic galileon terms. As a result, the degeneracy of the solutions which belong to the same families of the quadratic theory will be lifted, and they will be rearranged in {\it six} new families of solutions, in general:

\begin{itemize}
\item the family which extrapolated to regular Minkowski infinity will splinter into two new families. One will still behave as the parent 
family far from the source, with negligible cubic contributions, and $\varphi$ flowing to zero. In the other family the cubic contributions will remain comparable to the quadratic ones, such that they continue to nearly cancel far away, leaving a small mismatch that 
corresponds to the nonzero value of $Q$, as seen in the first of Eq. (\ref{system}). This family must have a larger $\varphi$ than the previous one, which therefore {\it cannot} flow to zero far away. Thus this family will be singular far away;
\item the doubly singular family will now split four ways. The reason is that it was characterized by different asymptotic limits
for $\varphi$ at the singularities $r=0$ and $r=\ell$, and $\varphi$ didn't vanish at either end. So, first the symmetry $\varphi \leftrightarrow - \varphi$ will be explicitly lifted, and furthermore the cubic terms on each individual branch will flow differently depending on the relative signs of the quadratic and cubic galileon contributions on the two branches. Hence, in the static limit we will see four different families of solutions, all singular at both ends, but interpolating between the two singularities in different ways. However, one of these families can be regulated far away if we allow it to become time-dependent, since it corresponds to the self-inflating galileon branch.
\end{itemize}

This is precisely confirmed by investigating (\ref{system}). From the 
first  and third of Eq. (\ref{system}), we can express 
$\nu'$ and $\sigma'$ as functions of $\varphi'$,
\ba
\nu' &=& -\frac23 e^{-2\nu-2\sigma}(r-q) + \frac{Q-e^{2\nu+2\sigma} \varphi'}{6\alpha \varphi'^2 e^{4\nu+2\sigma} }\, , \nonumber \\
\sigma' &=& \frac13 e^{-2\nu-2\sigma}(r-q) + \frac{Q-e^{2\nu+2\sigma} \varphi'}{6\alpha \varphi'^2 e^{4\nu+2\sigma} }
\, ,
\label{nusigma}
\ea
and then substituting these two in the second 
of (\ref{system}), after some straightforward rearrangements, obtain
\be
\frac{\left(Q-e^{2\nu+2\sigma} \varphi'\right)^2}{Q^2 + \frac23 (r-q)^2 + 2 e^{2\nu+2\sigma}} = 
\frac{6\alpha^2 \varphi'^4 e^{4\nu}}{1+6\alpha^2 \varphi'^4 e^{4\nu}} \, .
\label{mastereq}
\ee
This is a sextic in $\varphi'$, and for the general case of $Q \ne 0$ it has six different roots. These roots define the six
different branches of solutions of (\ref{system}), which are generated by taking any individual root of (\ref{mastereq}), substituting $\varphi'$ as a function of $\nu, \sigma$ and $r$ on it into (\ref{nusigma}), and integrating these two equations.
Note, that the expression for $\varphi'$ doesn't need to be integrated any further, because the actual value of $\varphi$ is not physically relevant, thanks to the pseudosymmetry of the theory (\ref{action}) defined by the simultaneous shift of $\varphi$ and the rescaling of dimensional parameters in the matter Lagrangian. 

We cannot solve (\ref{mastereq}) exactly, and write the specific roots in closed form. Fortunately, that is not necessary in order to understand how the solutions {\it behave}. We can determine the qualitative behavior of the solutions graphically, by considering the roots as the intersections of the two curves defined by the different sides of Eq. (\ref{mastereq}) at any fixed
value of $r$, and consequently of $\nu$ and $\sigma$, and then vary the radial location and the metric functions to see how the intersections behave. So, the left hand side (LHS) of (\ref{mastereq}) for fixed $r, \nu, \sigma$ is a parabola in $\phi'$, 
shifted from the origin by $Q e^{-2\nu - 2\sigma}$. The RHS defines a curve which is a quartic parabola near the origin, which changes to a slowly growing function saturating at unity as $\varphi' \rightarrow \pm \infty$. This function looks like a smoothed square well, because very close to the origin the quartic parabola is very shallow. In a generic situation, with $Q\ne0$, the two curves are offset relative to one another, so that they touch the horizontal axis at different locations. Therefore 
they must intersect at several points.

\begin{figure}[hbt]
\begin{center}
\includegraphics[width=0.8\textwidth]{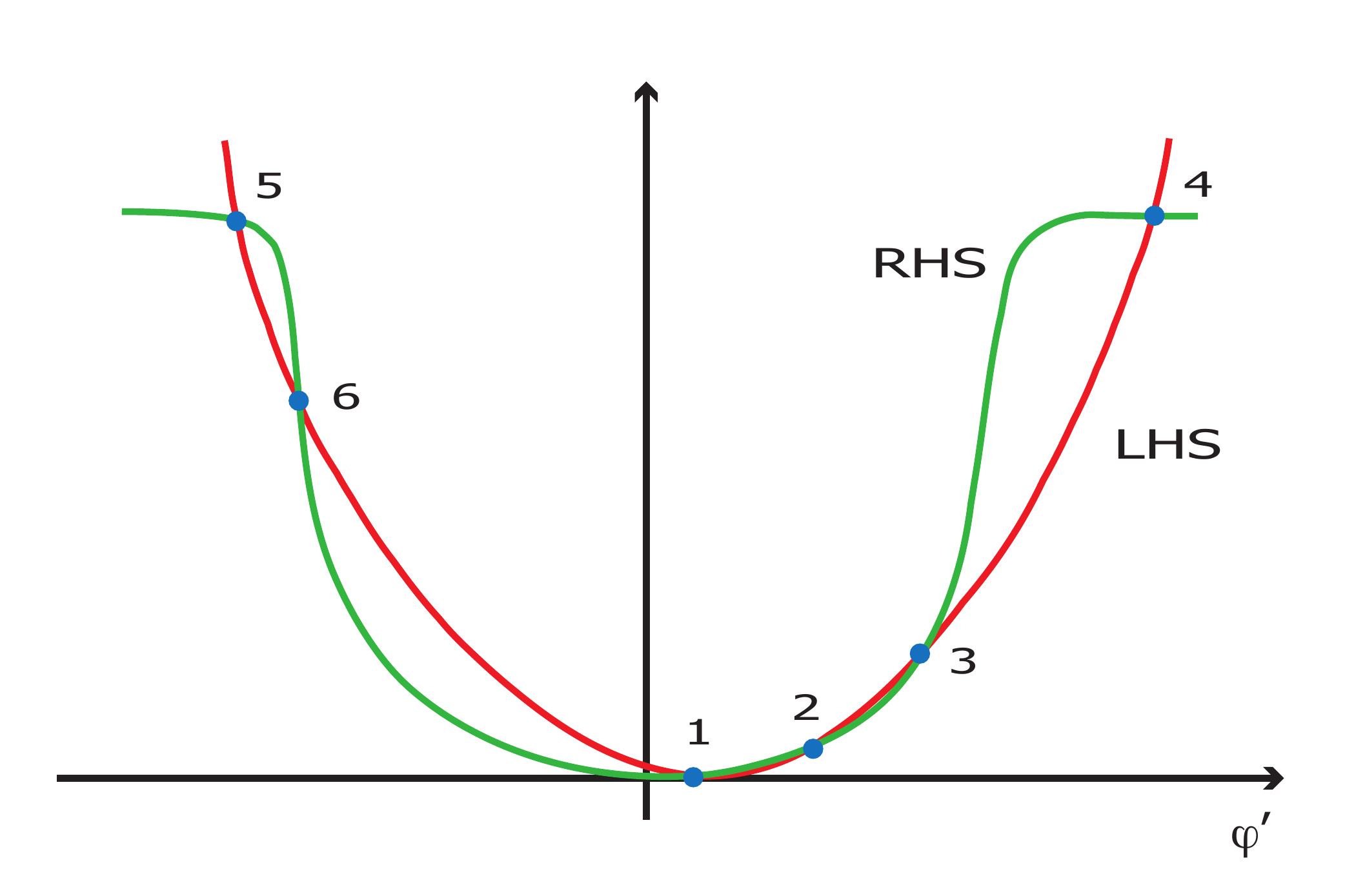}
\caption{Roots of Eq. (\ref{mastereq}) defining the branches of the cubic galileon theory.}
\label{curves}
\end{center}
\end{figure}

At radial distances close to a Dyson sphere, where the exterior metric coefficients are not too big - specifically, where the area of the angular sphere is not much larger than the area of the 
Dyson sphere itself - the two curves will intersect six times, as illustrated in Fig. (\ref{curves}). 
What happens is that $e^\nu = {\cal O}(1)$, because we are taking Dyson spheres to be larger than their Schwarzschild radius, so that their fields are not very strong, and near them the gravitational redshift is small. Simultaneously
$e^\sigma = {\cal O}(r)$, and so the coefficient of $\varphi'$ on the LHS is 
${\cal O}(e^{2\nu+2\sigma}) \simeq {\cal O}(r^2)$, and the shift from the 
origin is ${\cal O}(Q/r^2)$, being typically very small. On the other hand, while the quartic is very shallow
near the origin, its coefficient is ${\cal O}(\alpha^2)$, and because the length scale $\sqrt{\alpha}$ is very large, 
and so as $\varphi'$ increases, the RHS changes sharply and quickly saturates at unity. 

Now we can follow the system as the radial coordinate is increased, moving away from the Dyson sphere. Notice, that in this case the behavior of the system is very sensitive to which root we select, because the variation of the functions $\nu$ and 
$\sigma$ is controlled by (\ref{nusigma}). However: even without a detailed investigation, we can easily deduce that 
five of the six roots {\it must} be singular at some distance away from the Dyson sphere. Indeed, if we increase the functions $\nu$ and $\sigma$ in (\ref{mastereq}), the potential well shape in Fig. 
(\ref{curves}) narrows down, while the parabola moves towards the origin and becomes very steep very quickly, trying to
climb off of the potential well. 
Therefore the pairs of roots $3 \& 4$ and 
$5 \& 6$ are moving closer together, until they touch and eventually disappear altogether. This means that they {\it cannot}
exist after $\nu$ and $\sigma$ increase beyond some critical {\it finite} value, defined by the coincidence of the two roots. So,
in turn, $\nu$ and $\sigma$ cannot grow indefinitely if we are on one of these four roots.
Hence these four families of solutions {\it cannot} asymptote to flat Minkowski space, where $\sigma$ does grow beyond any bound. Furthermore, the moment when the roots osculate is a singularity of the system of differential equations, because the two pairs of families of solutions become degenerate. This means that the sphere of the radius where the degeneracy occurs must be a curvature singularity outside of the Dyson sphere, just like in the case of the doubly singular solutions in Brans-Dicke gravity (\ref{scalarsolns}). 
Note also, that we can estimate the critical distance where this occurs, by noting that the roots become degenerate when each side of the Eq. (\ref{mastereq}) is roughly ${\cal O}(1)$. This yields, with $e^\sigma \simeq r$, to $\varphi' \sim 1/r$ from LHS and $\varphi' \sim 1/\sqrt{\alpha}$ from RHS, and so $r_{cr} \sim \sqrt{\alpha}$.

On the other hand, roots $1$ and $2$ might allow $\nu$ and $\sigma$ to grow indefinitely on them, as $r$ grows larger.
To see what happens we can try to venture very far from the Dyson sphere on these two roots, and check if the fields are self-consistently getting weaker. 
To start, at very large $r$ we may try to approximate $\varphi' \ll 1/\sqrt{\alpha}$ and $e^\sigma \simeq r$ to the leading order, in order to check if these two roots are remaining close to the origin in Fig. (\ref{curves}) when $r, \nu, \sigma$ are growing. Substituting this approximation into  Eq. (\ref{mastereq}) yields 
$${\left(Q- e^{2\nu+2\sigma}\varphi'\right)^2}  =  4 \alpha^2 e^{4\nu+2\sigma} \varphi'^4 (1 + 3 e^{2\nu})$$
 where we have used the fact  that the denominator on the LHS of Eq. \ref{mastereq} goes like $2r^2/3+2e^{2\nu+2\sigma} \sim 2e^{2\sigma}/3+2e^{2\nu+2\sigma}$. Since to  leading order 
$\varphi' \simeq Q e^{-2\nu-2\sigma}$, setting $\epsilon = Q- e^{2\nu+2\sigma}\varphi'$ and plugging it here, we find
\be
\epsilon^2 \simeq 4 \alpha^2 Q^4 e^{-4\nu - 6 \sigma} \left(1+3 e^{2\nu}\right) \, .
\label{mastereqdec2}
\ee
So the roots $1 \& 2$ of the equation (\ref{mastereq}) are given by
\be
\varphi'_1 = Q e^{-2\nu-2\sigma} \Bigl(1 - 2 \alpha Q e^{-2\nu-3\sigma} 
(1+3e^{2\nu})^{1/2}\Bigr)  \, , ~~~
\varphi'_2 =  Q e^{-2\nu-2\sigma} \Bigl(1 + 2 \alpha Q e^{-2\nu - 3\sigma}
(1+3e^{2\nu})^{1/2} \Bigr) \, , \label{roots12}
\ee
which follows because the root $1$ is closer to the origin than the root $2$. Now we can substitute these two roots into 
Eq. (\ref{nusigma}), finding
\ba
\nu' &=& -\frac23 e^{-2\nu-2\sigma} (r-q) \pm 
\frac{1}{3} \left(1+3 e^{2\nu}\right)^{1/2} e^{-2\nu - \sigma} \, , \nonumber \\
\sigma' &=& \frac13 e^{-2\nu-2\sigma} (r-q) \pm 
\frac{1}{3} \left(1+3 e^{2\nu}\right)^{1/2} e^{-2\nu - \sigma} \, , \label{nusigm12}
\ea
where the top sign corresponds to the root 1 solutions, and the bottom sign to the root 2 ones, respectively. 

With the top sign, the function $\nu$ has a fixed point at infinity, where $\nu = 0$, and the terms ${\cal O}(1/r)$ in the top equation cancel precisely, leaving $\nu' = {\cal O}(1/r^2)$. Likewise, in this limit $\sigma' = 1/r + {\cal O}(1/r^2)$,
consistent with all of our approximations. The asymptotic behavior of the root 1 solutions at large distances is
\be
e^{2\nu} \rightarrow 1 - \frac {\cal \ell}{r} \, , ~~~~~~~~~~~   e^{\sigma} \rightarrow r \, ,
\label{root1asymptotics}
\ee
like the standard Schwarzschild solution and its scalar-tensor generalizations (\ref{scalarsolns}). 
The root 1 is certainly moving closer to the origin, and so it continues to exist.
Therefore, the branch defined by the root 1 is indeed asymptotically flat, and has fields 
which become very weak from the source. 

With the bottom sign, however, we find that sufficiently far away $\nu$ must stop growing and 
begin to decrease. The first of (\ref{nusigm12}) 
dictates that $\nu'$  must become negative. The key is that on this branch the scalar field gradient $\varphi'$ is slightly {\it greater} that $Q e^{-2\nu-2\sigma}$, and this is independent of the approximations made in getting Eqs. (\ref{nusigm12}). 
So, since near the Dyson sphere $e^\nu$ is positive, this implies it should vanish somewhere far. 
On the other hand, $\sigma$ may continue to grow, as controlled by the second of 
Eq. (\ref{nusigm12}). Since $\varphi' = Q e^{-2\nu - 2\sigma}$ to the leading order, on this branch the value of $\varphi'$ will grow, instead of decrease as on the asymptotically flat branch defined by root 1. So our approximation $\varphi' \ll 1/\sqrt{\alpha}$ will break down, and we need to go back to the full Eq. (\ref{mastereq}), where we should now consider the limit where $r, \sigma$ and $\varphi'$ are growing, while $\nu$ is decreasing, to determine what's really going on.
In this limit, the coefficient of the $\varphi'^2$ on the LHS of (\ref{mastereq}) is $e^{2\nu+2\sigma}/r^2$, and it is decreasing, while the location of the minimum, $Q e^{-2\nu - 2\sigma}$ is increasing. So the parabola is getting shallower and rapidly moving away from the origin, farther to the right. Meanwhile, on the RHS, since $e^\nu$ is decreasing, the potential well is getting wider. This means, that the roots $2$ and $3$ of Fig. (\ref{curves}) are moving {\it towards} one another as $r$ increases. At some distance they converge together, and become degenerate, which signals that they are singular there. So the solutions from the branch characterized by the root 2 will also have a naked singularity far from a Dyson sphere sourcing them, again at a distance $r_{cr} \sim \sqrt{\alpha}$.

It is instructive to consider various branches of the cubic galileon from the perspective of the decoupling limit, which is widely employed in galileon models, as well as other modified gravity setups. The decoupling limit is formally defined as the limit where gravity  is decoupled by taking $M_{Pl} \rightarrow \infty$, effectively fixing the background geometry, while keeping the scalar, its  self-interactions,
and any matter which sources the scalar. Since we have absorbed $M_{Pl}$ into $\varphi$ and $\alpha$, to enforce the decoupling limit in our equations we can merely rescale $\varphi \rightarrow Z \varphi$, $\alpha \rightarrow \alpha/Z$ and $Q \rightarrow Z Q$, take the limit $Z \rightarrow \infty$, fixing $e^{2\nu} = 1$ and $e^\sigma = r$. The equations in (\ref{system}) 
degenerate to a single scalar field equation,
\be
r^2 \varphi' + 4\alpha r \varphi'^2  = Q  \, ,\label{systemdec}
\ee
with only two roots,
\be
\varphi'_\pm  =\frac{r}{8\alpha} \left\{\pm  \sqrt{1 + 16 \alpha \frac{Q}{r^3} } - 1 \right\} \, .
\label{decroots}
\ee
These two roots define the branches of solutions given by roots 1 and 6 of the general case. 
Note that the normal behavior familiar from existing decoupling limit considerations require $Q \ge 0$. If $Q$ is negative, then the description based on (\ref{decroots}) breaks down at $r_* \simeq (\alpha |Q|)^{1/3}$, a scale which as we shall see shortly is just the Vainshtein scale of the source. 

The upper sign gives the root $\varphi' \rightarrow \frac{Q}{r^2}$ in the limit of large $r$. It corresponds to the
branch defined by the root 1 of the full theory, which admits a flat space limit, and interpolates between the near region of the Dyson sphere and Minkowski space at infinity. We stress again, that for the root 1 solutions to be really able to extrapolate between infinity and the surface of a small Dyson sphere with weak fields, the discriminant of the root equation must be positive, and $Q \ge 0$. If this is not the case, the Dyson sphere cannot develop large Vainshtein regions, as we will discuss in
more detail in what follows.

The lower sign  gives the root $\varphi' \rightarrow - \frac{r}{4\alpha}$, a confining solution, which yields the 
branch of solutions defined by the root 6 of the full theory. We can see this from (\ref{mastereq}), where in the decoupling limit the RHS reduces just to the quartic, implying that the roots 4 and 5 cannot be observed in the decoupling limit at all. Since the second root of (\ref{decroots}) is negative, it must be root 6. These solutions are really self-accelerating galileon backgrounds 
in disguise, as is clear from the decoupling limit\footnote{Note that if we tried to enforce the decoupling limit by `blunt force trauma' approach in (\ref{mastereq}) we would have found a square of (\ref{systemdec}) -- i.e. $r^2 \varphi' \pm 4\alpha r \varphi'^2  = Q$ -- admitting also roots 2 and 3. This is inconsistent, however, since these two branches do not exist outside of gravitational sector, as evidenced by their singularities. Technically, in obtaining (\ref{mastereq}) we have employed gravitational equations of motion, and encountered cancellations between powers of $M_{Pl}$, that yield spurions in the `decoupling limit' of (\ref{mastereq}). }. This is rather telling regarding the nature of the singularity at large distances on this branch. While it is certainly present for static backgrounds, since they have a confining potential far away, its appearance is a consequence of imposing the static limit. Consistent with this, the estimated location of the singularity at $r \sim \sqrt{\alpha}$ is precisely at cosmological scales where modifications of gravity alter the vacuum, and where $\sqrt{\alpha}$ plays the role of the Hubble scale. When galileon is allowed to be time-dependent, the geometry is completely nonsingular far away and in the future. Thus time dependence is a kind of `censor', removing the singularity on this branch. We will consider this in more detail in what follows. On the other hand, the asymptotically flat solutions are deformations of the flat vacuum induced by localized sources. Hence, it is those solutions which should be featuring the Vainshtein phenomenon, when we consider gravitational fields of Dyson spheres on that branch. We will examine it in detail in the next section. We also see that the decoupling limit analysis is blind to the existence of other, strongly gravitating solutions. This is not surprising, because we know that those solutions reduce to (\ref{scalarsolns}) in the limit $\alpha \rightarrow 0$, and those cannot be deduced without gravity. 

Finally, we note that this analysis confirms our assertion that the interior of Dyson spheres must be an empty space, because all the solutions which have nontrivial scalar hair have short distance singularities. As those are unacceptable for regular matter distributions, the interior must be free of scalar hair, and hence flat. 

\section{Vainshtein's Coiffure}

The branch defined by the root 1 of (\ref{mastereq}) yields solutions which are asymptotically flat. So, as noted above, this is the one candidate family of configurations which should feature Vainshtein mechanism. Let us examine them and find out  exactly what the Vainshtein mechanism {\it is}. In the literature it is often denoted by a term {\it Vainshtein's screening}. This is a misnomer, in our view. If we were to take the idea of `screening' literally, it would imply some effect of `renormalizing' the source charge similarly to what happens in electrolytes with Debye screening, or in quantum field theory. There, the external fields are compensated by the induced polarizations from real or virtual dipoles, arising in response to the external fields, and reducing the net charge of the source. Normally, the dipoles require negative charges, which is quite fine in gauge theory but possibly disastrous in gravity, where negative masses yield instabilities. On the other hand, gravity could be neutralized by `free fall', but here what needs to be suppressed is the forces due to the extra mode(s) (whose charge is still the mass, so the quandary of negative masses persists). 

The resolution is, as we find it, that in fact Vainshtein effect is really due to the dynamical redefinition of the long range hairs sourced by a mass rather than `conventional' screening. The net charge remains the same regardless of where it is measured, as controlled by the simple vacuum Gauss law. Instead, the field strength which it sources changed dynamically because of the irrelevant operators which become big closer in. Quantitative aspects of it, however, have been correctly anticipated in \cite{dvali}, which nevertheless assumed the decoupling limit. Here, we will in fact derive this in a self-consistent manner from the theory given by (\ref{system}), on the branch defined by root 1 of (\ref{mastereq}). This leaves no doubt as to the existence of smooth interpolating solutions across the Vainshtein's radius. 
Further, our construction reveals the limitations on the types of matter that feature the Vainshtein effect, revealing that the Vainshtein radius does depend on the specific microscopic details of the matter source, as anticipated in the formulation of the so-called {\it elephant problem} \cite{unpublished,tonyreview}. In this section, we will sketch the basic ideas which explain the Vainshtein effect, relegating the more precise, quantitative statements about matter featuring it to the next section.

So, let us start  by looking at the limit $\alpha = 0$, which is the pure Brans-Dicke theory, in order to define the problem. 
The extra scalar couples to matter through the conformal combination $\hat g_{\mu\nu} = g_{\mu\nu} e^{\beta \varphi}$. 
This is the cause of the problem. In perturbation theory, this gives the coupling ${\cal O}(1) \beta \varphi T^\mu{}_\mu$, in our normalizations. This means that a mass $M$ surrounds itself with the scalar hair. By the standard 
 $4D$ Gauss law, the scalar hair is
 \be
 \varphi' =  {\cal O}(1) \beta \frac{M}{M^2_{Pl} r^2} \, ,
 \label{schairs}
 \ee
 because the `charge' of the source is ${\cal O}(1) \beta M/M_{Pl}$. 
 Consequently, the energy density carried by the hair outside of the source is
 $\rho_\varphi \sim M_{Pl}^2 \varphi'^2 = {\cal O}(1) \beta^2 \frac{M^2}{M_{Pl}^2 r^4}$. In this conformal frame, the scalar energy density can't affect the gravitational potential of the source mass which is still given by the Newtonian expression $V_N = - \frac{G_N M}{r}$. However, probes couple directly to the scalar hair, because they move on geodesic trajectories of the conformal metric $\hat g_{\mu\nu}$, and not of $g_{\mu\nu}$. The matter trajectories are defined by
\be
\frac{d^2x^\mu}{ds^2} + \hat \Gamma^\mu_{\nu\lambda} \frac{dx^\nu}{ds} \frac{dx^\lambda}{ds} = 0 \, , 
\label{geodesic}
\ee
and so the force from a spherical mass distribution 
which acts on non-relativistic matter probes is $F_{\rm net} = \hat \Gamma^r_{00}$. 
In terms of the original metric $g_{\mu\nu}$,
using the standard conformal transformation formula with the conformal factor $\Omega^2 = e^{\beta \varphi}$, this gives
$F_{\rm net} = \Gamma^r_{00} - \frac12 \frac{\partial_r \Omega^2}{\Omega^2}$. Then, using 
$\Gamma^r_{00} \sim - \frac12 \partial_r h_{00} \sim - \partial_r V_N = - G_N M/r^2$
and $\partial_r \Omega^2/\Omega^2 \sim \beta \varphi' =  {\cal O}(1) \beta^2 M/(M_Pl^2 r^2) 
=  {\cal O}(1) \beta^2 G_N M/ r^2$, we find that the net force is
\be
F_{\rm net} = -G_N \frac{M}{r^2} (1+ {\cal O}(1) \beta^2) \, .
\label{force}
\ee
If the effective coupling $\beta^2$ is not smaller than about $5 \times 10^{-3}$, 
the theory would run into problems with solar system bounds, coming from the Cassini satellite.

Turn $\alpha$ back on. The Gauss law formula (\ref{schairs}) changes to 
\be
\varphi' + 2 \alpha (\nu' + 2 \sigma') \varphi'^2 e^{\nu} =  {\cal O}(1) \beta \frac{M}{M^2_{Pl} } e^{-2\nu - 2\sigma} \, ,
\label{newgauss}
\ee
as per the first of Eq. (\ref{system}). Here we have used 
$Q = \beta(\rho-2p) r_0^2/2 = {\cal O}(1) \beta M/M_{Pl}^2$\footnote{Here, ${\cal O}(1)$ term disguises the sensitivity of the hair to the type of matter distribution, i.e. the elephant problem, alluded to by the $\rho-2p \sim 1-2w$ prefactor, which we will discuss more in the next section.}, found after inserting the first of Eq. (\ref{bcs}) into the first of Eq. (\ref{system}). We take $Q \ge 0$, necessary to interpolate weak fields to short distances, as noted previously\footnote{We will also discuss this in more detail in the next section.}. For Dyson spheres larger than their Schwarzschild radius, $r_0 > r_s \sim G_N M$, on this branch we can consistently approximate $e^{2\nu} \simeq 1 - \frac{\ell}{r}$ and $e^\sigma \simeq r$. 

\begin{figure}[hbt]
\begin{center}
\includegraphics[width=0.9\textwidth]{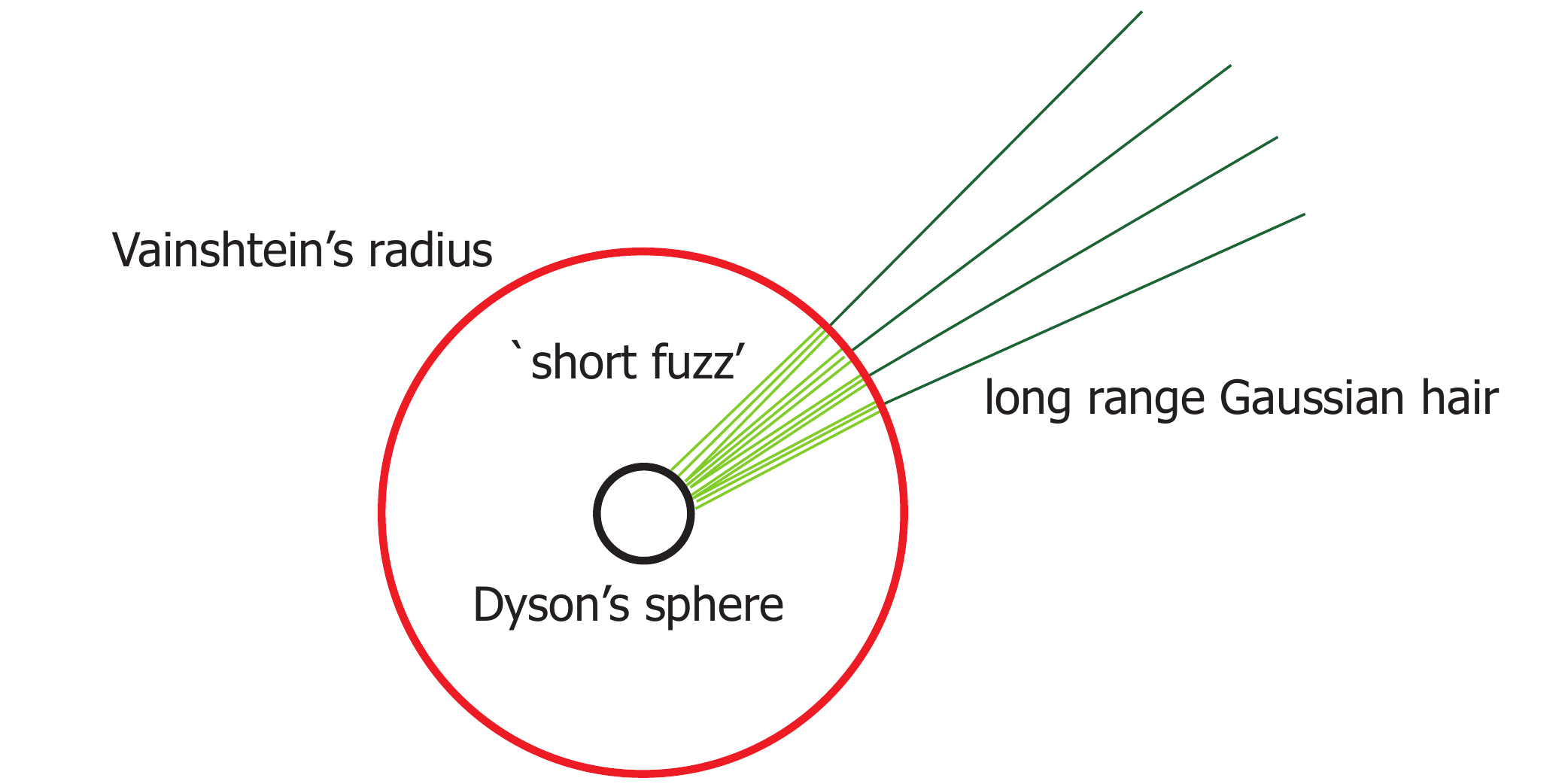}
\caption{Vainshtein's coiffure.}
\label{vainpic}
\end{center}
\end{figure}

So to the leading order we find precisely the decoupling limit equation for the scalar,
\be
\varphi' + 2 \alpha \frac{\det g'}{\det g} \varphi'^2 =  {\cal O}(1) \beta \frac{M}{M^2_{Pl}  r^2} \, ,
\label{decls}
\ee
where we have rewritten $2\sigma' = \frac{2}{r} = \frac{\det g'}{\det g}$ in a more illustrative way, as the logarithmic derivative of the metric determinant. This changes how the field lines distribute over a fixed solid angle at shorter distances. Indeed, {\it if the configuration is asymptotically flat}, at very large distances the Gauss law fixes the field line source to be $\sim M/r^2$. There asymptotic flatness ascertains that the cubic terms are tiny and the hair (\ref{decls}) is approximately equal to (\ref{schairs}). 
As the distance decreases, the hair grows thicker, the cubic terms catch up, and begin to compete with the standard 
$\varphi'$ field strength. This occurs at $\varphi' \sim \alpha \frac{\det g'}{\det g} \varphi'^2$, or therefore where 
$\varphi' \simeq {\cal O}(1) \beta \frac{M}{M^2_{Pl}  r^2} \simeq r/\alpha$. At shorter distances, the cubic contributions completely take over. 
The transition scale is precisely
\be
r_V \simeq \left(\frac{M \alpha |\beta|}{M_{Pl}^2}\right)^{1/3} \simeq \left(r_S \alpha |\beta|\right)^{1/3} \, ,
\label{vainshscale}
\ee
i.e. it is the Vainshtein's scale\footnote{Actually, it depends on the type of matter source, being proportional to $\rho-2p \sim 1-2w$.}. Below $r_V$ this yields a different scaling for $\varphi'$; an additional effect comes from $\det g'/\det g \simeq 2/r$ in Eq. (\ref{decls}), which dramatically slows down the bunching of field lines at shorter distances. This term partially closes the `angular' $2$-sphere to the galileon field lines, which therefore have a larger spread over a given solid angle at shorter distances. Taking Eq. (\ref{decls})
into account for Dyson spheres which are much larger than their Schwarzschild radius, at distances below $r_V$, we find
\be
\varphi' = {\cal O}(1) \left( \frac{|\beta| M}{\alpha M_{Pl}^2 r} \right)^{1/2} = {\cal O}(1) \left( \frac{|\beta| r_S}{\alpha r} \right)^{1/2}  = {\cal O}(1) \beta \,  \frac{r_S}{r_V^2} \,  \sqrt{\frac{r_V}{r} } \, .
\label{fuzz}
\ee
The field lines are {\it less dense} at short distances than they would have been without the galileon contributions. 
Their distribution is depicted in Fig. (\ref{vainpic}). The galileon suppression by the higher dimension operator involving the derivative of the volume determinant makes
them look more like a fuzz, spreading them out and preventing `hair bunching' closer to the source. 
Further, because the length scale $\sqrt{\alpha}$ is taken to be large, the transition scale for this `fraying', or `splintering' of the scalar hair $r_V$ is very large compared to the gravitational radius of the source. Hence where the transition occurs the field strength is very weak, and so it remains 
 very weak all the way down to the source, thanks to the modified distribution of field lines (\ref{fuzz}).

This suppresses the galileon corrections to forces acting on probes inside the Vainshtein radius. 
First, they do not gravitate very much. The scalar energy density is 
$\rho \sim \alpha \frac{\det g'}{ \det g} \varphi'^3 \sim \alpha \beta^{2} \frac{r_S^3}{r_V^7} (\frac{r_V}{r})^{5/2}$. 
Because $r_S < r \ll r_V$,  $\rho \le \beta^{2} (\frac{r_S}{r_V})^{1/2} \frac{1}{r_V^4}$, this is much too small to 
directly alter the background geometry, as can be readily checked by comparing it to the RHS of the second of 
Eq. (\ref{system}). Since the nontrivial gravitational corrections scale as $\sim r_S/r^3$, they are orders of magnitude larger
than the galileon sources, shifting the metric by terms $\propto 1/\sqrt{r}$, which dilute more slowly than 
$1/r$, but are {\it tiny} to start with. Second, and as before, since matter moves in the background given by $\hat g_{\mu\nu} = e^{\beta \varphi} g_{\mu\nu}$,  the effective force a particle following the geodesics (\ref{geodesic}) is $F_{\rm net} = \Gamma^r_{00} - \frac12 \frac{\partial_r \Omega^2}{\Omega^2}$. The conformal factor derivative now is still $\propto \varphi'$,
but the scalar gradients are different. From (\ref{fuzz}) we find
$\partial_r \Omega^2/\Omega^2 \sim \beta \varphi' =  {\cal O}(1) \beta^{2} \,  \frac{r_S}{r_V^2} \,  \sqrt{\frac{r_V}{r} }$. Therefore the net force acting on nonrelativistic probes at distances below the Vainshtein's scale is
\be
F_{\rm net} = -G_N \frac{M}{r^2} \left(1+ {\cal O}(1) \beta^{2} \left(\frac{r}{r_V}\right)^{3/2}\right) \, .
\label{forcevain}
\ee
Integrating this once, the net nonrelativistic gravitational potential a particle inside the Vainshtein radius is subject to is
\be
V = -G_N \frac{M}{r} \left(1 +  {\cal O}(1) \beta^{2} \left(\frac{r}{r_V}\right)^{3/2} \right) \, . 
\label{potvain}
\ee
The correction is $\Delta V/V_N = \beta^{2} (\frac{r}{r_V})^{3/2}$, and, leaving aside the galileon coupling  
$\beta$ and the type of matter correction $ \propto 1-2w$, this is precisely what was found in DGP model, in \cite{gruzinov}. This confirms that the cubic galileon model can be viewed as a decoupling limit of the simplest $5D$ codimension-1 setup in 
DGP \cite{dgp}, on normal branch where ghosts do not roam \cite{tony&co}. The consequences of the correction are small, as noted in \cite{gruzinov,glenn}, with the dominant effect being the precession of bound orbits, as per the (violation of) the Bertrand's theorem, which implies that any deviations of central forces that differ from $\sim 1/r^{2}$ or $\sim r$ yield orbits that are not closed. 
This has been argued to be relatively small, but only by an order of magnitude or so below the current experimental sensitivity of lunar laser ranging \cite{giazalgru}. In fact, for the case of galileon, the coupling parameter $\beta$ gives a better chance for observability, and the simple cubic galileon may already be constrained by the lunar laser ranging, as we shall discuss below.

The analysis above explains the argument that the Vainshtein effect is a consequence of long range hair fraying rather than `conventional' screening as in gauge theories. The total charge, defined by the usual $4D$ Gauss law never changes, but the field lines which it sources are redistributed between the extra field strengths included in the irrelevant operators that take over at sub-Vainshtein distances. 

In closing our {\it expos\'{e}} of the Vainshtein effect, let us note its implications for the perturbative description of the scalar fluctuations around the background. The key is to determine the full quantum cutoff, which as noted in \cite{lpr} is the scale beyond which the gravitational effects are probing the trans-Planckian distances. Since the scalar perturbation theory naively breaks down at the scale where the cubic scalar terms in (\ref{action}) take over from the quadratic ones. This occurs at $1/\hat \alpha^{1/3}$ \cite{lpr}, or equivalently at the distance scale $r_{\rm cutoff} \simeq ({\ell_{Pl} \, \alpha})^{1/3}$ (which is parametrically the Vainshtein radius of a Planckian black hole). 
If we were to take $\sqrt{\alpha}$ to be Hubble's distance, $\sqrt{\alpha} \simeq H_0^{-1} \simeq 10^{61} \ell_{Pl}$, so as to ensure that  it is only  cosmological solutions and little else that are affected by the modification of gravity, this would have yielded the $r_{\rm cutoff} \simeq 10^{41}  \ell_{Pl}$,
or therefore $r_{\rm cutoff} \simeq 10^{3} ~{\rm km}$. Such a low scale is the main source of problems in DGP, as noted by \cite{lpr}.

However, what really breaks down at this scale is the {\it linearized description} of scalar perturbations around the 
the far-field vacuum, at super-Vainshtein scale distances. Below the Vainshtein scale one needs to account for the background effects of the scalar hair which gives larger contribution to the scalar propagator than the vacuum terms. This renormalizes the scalar perturbation kinetic term by a factor $Z \simeq \alpha \varphi'' \simeq \alpha \varphi'/r$, which is large for $r < r_V$, and pushes down the galileon self-coupling, given by the cubic scalar term in 
 Eq. (\ref{action}). After canonically normalizing the fields, this renormalizes the canonical coupling $\hat \alpha \rightarrow \hat \alpha/Z^{3/2}$.
\cite{ratnic,tanaka}. The scalar cutoff is the energy scale below which the renormalized cubic is smaller than the quadratic, and so it is given by
$\Lambda_{\rm cutoff} \simeq (Z^{3/2}/\hat \alpha)^{1/3} \simeq Z^{1/2} (M_{Pl}/\alpha)^{1/3}$. So the short distance cutoff $r_{\rm cutoff} \simeq 1/\Lambda_{\rm cutoff}$ is $r_{\rm cutoff} \simeq (\ell_{Pl} \alpha)^{1/3}/Z^{1/2}$. 
Using (\ref{vainshscale}) and (\ref{fuzz}), the suppression factor is $Z \simeq \left(\frac{r_V}{r} \right)^{3/2}$, where $r_V$ and $r$ are the Vainshtein radius of the source of the galileon hair, and $r$ the distance where the hair strength is evaluated, respectively.
Thus, at a distance $r$ from the center of a source of mass $M$, the effective cutoff is
\be
r_{\rm cutoff} \simeq (\ell_{Pl} \alpha)^{1/3} \, \Bigl(\frac{r}{r_V} \Bigr)^{3/4} = 
\Bigl(\frac{\alpha}{\ell_{Pl}^2}\Bigr)^{1/12} \, \Bigl(\frac{r^3}{|\beta| M}\Bigr)^{1/4} \, .
\label{cutoffr}
\ee
In terrestrial conditions,  with $\sqrt{\alpha} \sim H_0^{-1} \sim 10^{61} \ell_{Pl}$, this gives  
\be
r_{\rm cutoff} \simeq  \frac{1 {\rm cm}}{~|\beta|^{1/4}} \, ,
\label{cutoffUV}
\ee
of order the Earth's Schwarzschild radius. 
This simple analysis generalizes the revised DGP cutoff given in \cite{ratnic}. While this cutoff is very low, a price to pay for having a large length scale $\sqrt{\alpha} \sim H_0^{-1}$, which yields large Vainshtein's radii and efficient suppression of extra forces at distances shorter than them, while having a chance at yielding branches where cosmological effects of the galileon may play a role,  it can be pushed below the current experimental bounds, at around ${\cal O}(100)$ microns in the gravitational sector \cite{eotwas}, by increasing the coupling to matter $|\beta|$. Indeed, it is clear from (\ref{forcevain}) that we can increase $|\beta|$ by a couple of orders of magnitude without affecting the force significantly in the laboratory conditions, because the Vainshtein radius can be kept large and the hair near the source thin. Note that such increase of the extra force would not affect weak equivalence principle bounds, since it is independent of the type of matter probes. On the other hand, it may also happen that even though the galileon sector ceases to be predictive below a centimeter,
whatever resolves it cannot produce sufficiently dramatic effects to influence any terrestrial experiments at sub-millimeter scale, thanks to the huge suppression generated by the large Vainshtein radius of the Earth, 
$({\rm cm}/{r_{V \oplus}})^{3/2} \sim 10^{-28}$, as argued by \cite{ratnic}. 
Thus one could take a view that while galileon models can't make any observable predictions for tabletop searches for submillimeter deviations from Newton's gravity, they may at least be safe from any bounds that follow from tabletop experiments. 

So, ignoring theoretical prejudices, this shows that in galileon models with more tunable parameters controlling the interactions may at least give a chance for searches of their signatures by correlating cosmological and astrophysical searches. As we noted above, lunar laser ranging is already constraining cubic galileon theory, because the precession of lunar perigee here may be faster than in DGP. The precession rate is controlled by the relative correction to the Newtonian potential \cite{giazalgru}, and in our case the advance per orbit is
\be
\delta \phi = \frac{3\pi}{4} \frac{\Delta V}{V_N} = |\beta|^{3/2} \delta \phi_{DGP} \, ,
\label{precession}
\ee
which means that increasing the matter coupling $\beta$ by only a factor of a few, while keeping other parameters unchanged, brings the cubic galileon effects to observable levels. This yields a constraint 
$\beta \le {\cal O}(10)$ if $\sqrt{\alpha} \sim H_0^{-1}$. It would be interesting to see a more thorough investigation of possible galileon signatures.

\section{Combing the Coiffure: Boundary Conditions}

So far we have been discussing the solutions ``in the bulk", branch by branch, in order to gain insight in the qualitative behavior of geometries corrected by galileon hairs. Here we complete the discussion by showing how to construct the realistic fields of Dyson spheres by using the junction conditions (\ref{bcs}) to connect the interior and the exterior into the complete configuration. After we take 
the interior of the Dyson sphere to be a ball of Minkowski space, we must pick the same coordinate gauge as the exterior metric, and further ensure that in this gauge the Dyson sphere sits at the same coordinate location $r_0$ as measured from the outside.  

So, starting with $ds^2_- = - d\hat t^2 + d \hat r^2 + \hat r^2 d\Omega_2$, and recalling that in our gauge the exterior metric is $ds^2_+ = - e^{2\nu} dt^2 + e^{-2\nu} dr^2 + e^{2\sigma} d\Omega_2$, we see that the appropriate coordinate transformation in the interior is $\hat t = e^{\nu_0} t$, $\hat r = e^{-\nu_0} [r + (e^{\nu_0} - 1) r_0]$, where $r_0 = e^{\sigma_0}$, and the subscript zero implies that the functions bearing it are evaluated at $r=r_0$. Physically, this coordinate transformation simply means that we have to synchronize the clocks just inside and outside of the Dyson sphere, accounting for the gravitational redshift at the sphere relative to an observer at infinity. Following this, however, we must shift the radial coordinate to compensate for the Lorentz contraction of the radial coordinate in our gauge, induced by the redshift. This gives 
\be
ds_-^2 = - e^{2\nu_0} dt^2 + e^{-2\nu_0} dr^2 + e^{-2 \nu_0} \left(r+(e^{\nu_0}-1)r_0 \right)^2 d\Omega_2 \, 
\label{interior}
\ee
for the interior metric. Therefore, on the shell we have
\be
e^{\nu} = e^{\nu_0} \, , ~~~~~ e^{\sigma} = r_0 \, , ~~~~~  \nu'_- = \varphi'_- = 0 \, , ~~~~~  
\sigma'_{- } = \frac{e^{-\nu_0}}{r_0} \, .
\label{shells}
\ee
Substituting this in the junction conditions (\ref{bcs}), and subtracting the second from the third of (\ref{bcs}), yields the derivatives just outside,
\ba
&&\varphi'_+ + 2\alpha e^{2\nu_0} \left( \nu'_+ + 2\sigma'_+ \right) \varphi'^2_+ 
= \frac{\beta}{2} e^{-2\nu_0} (\rho - 2p) \, , \nonumber \\
&& \sigma'_+ = \frac{e^{-\nu_0}}{r_0} - e^{-2\nu_0}\frac{\rho}{2} + \frac{\alpha}{3} e^{2\nu_0}\varphi'^3_+  \, , \nonumber \\
&&\nu'_+ = e^{-2\nu_0} \frac{\rho + 2p}{2} + \frac{\alpha}{3} e^{2\nu_0} \varphi'^3_+  \, .
\label{bcsout}
\ea
Combining the first of these equations with the first of (\ref{system}), and the difference of the third and second 
with the third of (\ref{system}) yields,
respectively,
\be
Q = \frac{\beta}{2} (\rho - 2p) r_0^2 \, , ~~~~~~~~~~~
q = (\rho + p) r_0^2 + (1-e^{\nu_0}) r_0 \, .
\label{intcs}
\ee
The first of these two relations has been used in the previous section already. 

The second of Eq. (\ref{bcsout}) is the condition which fixes the radius of the Dyson sphere, once 
its energy density and pressure, and the branch of the exterior solutions are given. This equation may have several roots, or none, depending on the branch and $\rho$ and $p$. The second 
equation of (\ref{system}), when evaluated at the Dyson sphere,
\be
\varphi'^2_+ + 4\alpha (\nu'_+ + 2\sigma'_+) \varphi'^3_+ e^{2\nu_0} = 2(2 \nu'_+ \sigma'_+ + \sigma'^2_+) - 
\frac{2 e^{-2\nu_0}}{r_0^2} \, . 
\label{radius}
\ee
does not yield further significant constraints on the parameters of the solution. When $\varphi'=0$, its RHS has a nontrivial eigensolution $e^{2\nu} = 1 - \ell/r, e^{\sigma} =r$, on which it vanishes.  This is just the Schwarzschild solution, and we could always cut it at an arbitrary radius after specifying the right energy density and pressure at the shell. These many degenerate solutions mean that this equation merely controls the deviation of $\nu'$ away from the Schwarzschild form as governed 
by the scalar sources, rather than fixing the radius of the Dyson sphere.
In the weak field limit, we can therefore ignore it as redundant. 

So now in principle one can work out the exterior derivatives for any of the bulk branches and find the full exterior geometry.
Without galileon modifications, we normally keep the solutions which asymptote to flat space as the physically relevant ones, and throw away the ones singular far away as having unacceptable boundary behavior. Hence the most pressing check is to see whether the galileon terms alter this in some way. 

On the one hand, as we have already indicated previously, when $\alpha >0$, the sources with $\beta (\rho - 2p)<0$ prevent weak fields from setting up inside the Vainshtein sphere.  To see it, it suffices to consider the decoupling limit, where the root 1 branch is determined by the upper sign solution of Eq. (\ref{decroots}). Using the first of
(\ref{intcs}) yields
\be
\varphi'_+  =\frac{r}{8\alpha} \left\{\sqrt{1 + 8 \alpha \frac{\beta (\rho-2p)r_0^2}{ r^3} } - 1 \right\} \, .
\label{rootone}
\ee
When  $\alpha \beta (\rho - 2p)<0$ the discriminant vanishes precisely at $r \sim (\alpha |\beta (\rho - 2p)| r_0^2)^{1/3} \sim r_V$. 
If we define the shell equation of state, $p=w\rho$, this gives $r_V =  (\alpha |\beta (1- 2w)| \rho r_0^2)^{1/3}$. The weak field description below the Vainshtein radius fails, and if the discriminant vanished for any finite value of $r$, by continuity we would expect a branch merging where uniqueness of solutions breaks down, which signals a singularity. So a Dyson sphere would have to be placed farther out from any such surface, when the stress energy on it satisfies the condition $\alpha\beta (1-2w)<0$.

On the other hand, when the sources have stress energy satisfying $ \rho <-2p$, the shells cannot be static. To see this, we can treat the system (\ref{bcsout}) perturbatively, ignoring the scalar field at first. In that case, the regular exterior solution reduces to Schwarzschild, $e^{2\nu} = 1 - \ell/r$ and $e^\sigma = r$. The boundary conditions (\ref{bcsout}) then give 
\be
\ell = (\rho + 2p) r_0^2 \, , ~~~~~~~~~ \rho = \frac{2}{r_0} \sqrt{1- \frac{\ell}{r_0}} \,\, \left( \,1 - \sqrt{1-\frac{\ell}{r_0}}  \,\, \right) \,. 
\label{schbcs}
\ee
Note, that the first formula reproduces the correct expression for the active gravitational mass of a source \cite{shang}. Clearly, if we take $\rho + 2p<0$, $\ell$ must be negative by this equation.  But then the second equation can only have a negative solution for energy density.  For sources obeying null 
energy condition, $\rho + p  \ge 0$,
this requires $p \ge - \rho = |\rho|$, yielding $\rho + 2p >0$, which contradicts the starting assumption.
Thus, barring direct null energy violations in the matter sector, as we claimed we must have $\rho > -2p$, yielding $\ell > 0$, for a static shell to have any energy on it at all. In fact, it is easy to understand this intuitively. The point is, that if we were able to make a static Dyson sphere with sources satisfying $\rho + 2p <0$, we would have obtained an exterior solution which would look like Schwarzschild with negative mass, and without singularities in the core. That would have yielded many problems in the theory, starting with instabilities and so on \cite{myershorowitz}. Hence, such solutions should not be possible, and indeed the junction conditions
exclude them. 

When we turn on the scalar, the conclusions do not change for asymptotically flat solutions, 
$r \ge \ell$, if we ignore the galileon terms, because the junction conditions for the metric remain qualitatively the same
after we substitute (\ref{scalarsolns}) into (\ref{bcsout}). Once we turn on the galileon corrections, the extra terms on the RHS 
of the second of Eq. (\ref{bcsout}) could, for $\varphi'>0$, translate into additional positive contributions on the RHS of
the second of Eq. (\ref{schbcs}). One may worry that such corrections could yield static Dyson spheres with positive energy density, but obeying $\rho + 2p <0$. 
However, this is very difficult and technically requires huge Dyson spheres, that may be very unstable and subject to large IR corrections. We can see this as follows.
Suppose that 
$\alpha$ is very large, so that Vainshtein effect may be operable. It makes the extra galileon corrections 
tiny, $\varphi' \le 1/\sqrt{\alpha}$. So, any solution with $\rho > 0$ and $\rho + 2p<0$ must have at most a tiny energy density 
$\rho \sim 1/\sqrt{\alpha}$. When the matter does not violate the null energy condition, the inequalities $\rho + 2p <0$ and $\rho + p>0$ further limit the pressure, $p \sim {\cal O}(\rho)$. So this gives 
$\ell \sim {\cal O}(r_0^2/\sqrt{\alpha})$,  and that, from the second of Eq. (\ref{schbcs}) with the galileon corrections included, would yield  $r_0 \geq {\cal O}(\sqrt{\alpha})$, making the Dyson sphere larger than
the scale where gravity is modified! In other words, such spheres would have to be cosmologically large. So in a cosmological setting, the cosmological effects, and causality in the very least would obstruct their 
existence\footnote{Nonetheless, it is curious that at least in theory one might imagine constructing such huge negative mass static sources, even on a branch of the theory where there is no ghost. 
In fact, one may use the existence of such solutions to place a bound on $\sqrt{\alpha}$. If one wishes to leave no chance for any late formation of such large spheres with repulsive Newtonian potentials, and prevent any chance of them wreaking havoc in a late universe, one concludes that $\sqrt{\alpha} \ga H_0^{-1}$. }.

As a bottomline, we see that the `conventional' Vainshtein effect may be operable for sources on the Dyson sphere which satisfy the following energy conditions
\be \label{energyconds}
\alpha \beta (\rho- 2p) >0,  \qquad \rho+2p>0, \qquad \rho >0 \, .
\ee
For these sources, static Dyson spheres may be built, and their external fields are to leading order conforming to general relativistic expressions at distances below the Vainshtein radius. At larger distances, the fields reduce to the linearized  solutions with scalar fields (\ref{scalarsolns}), behaving as in a scalar-tensor theory. So, for sufficiently large $\alpha$, giving a large Vainshtein radius, the theory may be in accord with observational bounds, and the scalar forces may indeed be negligible, if we take solutions on the branch of root 1.  In order to recover the results of conventional General Relativity, specifically for sources which behave as nearly pressureless dust, from this equation we see that $\alpha \beta$ should be positive. 
This means, that the sources with more negative pressure cannot benefit from the protection of the Vainshtein effect, and the fields of extra modes will not stay small near such sources. Hence arguments that the cosmological volumes of matter must automatically be subject to the description appropriate to the `interior' of the Vainstein sphere \cite{dvali} may be incompatible with the operability of the Vainshtein effect at shorter scales if the net effective pressure inside the cosmological region is sufficiently negative to reproduce cosmic acceleration.

Note, that the Vainshtein radius depends not only on the total mass of the source, but also on the type of matter which is sourcing the fields. For  matter satisfying (\ref{energyconds}) the Vainshtein radius is given more precisely by
\be
r_V \sim \left(\frac{1-2w}{1+2w} \alpha \beta r_S \right)^{1/3} \, ,
\label{preciserv}
\ee
which shows that the suppression of extra forces only works when pressures are not too large. As the pressure increases, the
Vainshtein radius shrinks, while simultaneously the scalar hair, and the force it generates, became smaller. Once the scalar gradients become negative, however, the suppression of the hair is impossible for asymptotically flat solutions, since the
two terms in the galileon conservation law get locked together, and so the hair does not splinter enough.
As we noted, this supports the idea that the Vainshtein effect must be sensitive on the details of the source which generates the fields rather than just its total mass, as suggested by the elephant problem.

\section{Hirsute Bubbles: Branch Changelings}

The discussion of the previous section shows that Dyson spheres which carry stress energy with $0< \rho < -2p$ won't remain static. 
That is as it should be, since generically such shells are spherical domain walls separating different vacua of the matter sector,
and will eventually expand out, generically becoming null at late times, with $\rho \rightarrow -p$. In conventional General Relativity, these domain walls separate bubbles of lower vacuum energy from the parent vacuum with higher energy density, 
and are widely used for cosmological applications (see, e.g. \cite{bubbles}). In modified gravity models with multiple branches, a
new possibility arises, where bubbles may catalyze transitions between branches. Here we shall outline this for the cubic galileon model. Nevertheless we think this possibility is fairly generic, and needs to be understood in more detail.

In what follows, we will make several simplifying assumptions, since we only aim to demonstrate that branch transitions are possible in principle. For this reason, we will work with the assumption that the vacuum energy on each side of the bubble is the same, and that the energy difference between the sides is provided by the galileon gradients. Further, we will work perturbatively around  the decoupling limit, which simplifies the calculation, and at least allows us to consistently describe the asymptotically flat and 
self-accelerating branches. The latter assumption in itself is not too restrictive for large bubbles, but the former one is more special, and in general one would like to extend the analysis by adding generic nonvanishing cosmological terms off the shell.
We hope to return to this elsewhere. 

Bubble mediated branch changes have been considered in the cubic galileon in the decoupling limit previously, in \cite{koyama-tanaka-pujolas}. These authors found an instanton describing the nucleation of a tensionless bubble separating two branches, but then they argued that such solutions were merely a quirk of the thin wall limit, and that tunnelling between branches was in fact prohibited. Their point was that the vanishing effective tension was a sign of exact cancellation of
positive and negative energy densities, where the latter had to be put in by hand to mediate the transition, and was otherwise extraneous to the theory. We remain dubious about this interpretation. The self-accelerating branch contains a perturbative ghost \cite{tony&co}, and generically one expects such branches to be always present in multi-branch models \cite{bigal}. So such ghosts may provide the requisite negative energy in the core of the wall, with a perfectly physical  matter sector contributing positive energy. Obviously, this implies that the self-accelerating side of the bubble must be modified relative to the geometry used in \cite{koyama-tanaka-pujolas}, but that may be necessary anyway once the ghost is accounted for. Motivated by this reasoning, we will reconsider the issue of branch changing bubbles here, and, in the leading order gravity approximation, provide solutions that describe nucleation of hirsute bubbles with large and positive tension, which circumvent the no-go claims of \cite{koyama-tanaka-pujolas}.

Let us now show how to construct such branch changelings in the theory (\ref{action}). We take the parent state to be a vacuum, either on the asymptotically flat branch or on the self-accelerating branch, and the descendant to be the state with a bubble of the other vacuum surrounded by the parent vacuum. For simplicity, we will set vacuum energy contributions from the matter sector to be zero, and work only with the self-accelerating contributions from the galileon. This suffices for our purposes here. To describe a bubble, we take an $SO(3, 1)$ symmetric configuration
\be \label{so31}
ds^2=g_{\mu\nu}dx^\mu dx^\nu=e^{2\pi(\zeta)}\left( d\zeta^2+\zeta^2 \gamma_{ij} dx^i dx^j\right) \, , 
~~~~~~~~~ \varphi=\varphi(\zeta) \, ,
\ee
where $\gamma_{ij} dx^i dx^j$ is the metric on three-dimensional de Sitter space with unit curvature. 
We take the bubble wall to lie at a fixed ``radius", $\zeta=\zeta_w>0$, such that the  field profiles are given by
\be
\pi(\zeta)=\cases{\pi^{in}(\zeta) & $\zeta<\zeta_w$ \cr \pi^{out}(\zeta) & $\zeta>\zeta_w$ } \, , \qquad \varphi(\zeta)=\cases{ \varphi^{in}(\zeta) & $\zeta<\zeta_w$ \cr \varphi^{out}(\zeta) & $\zeta>\zeta_w$ } \, .
\ee
The interior and exterior are taken to lie on opposite branches. 
This ansatz is just a deformation of the standard VIS geometry used to describe tensional domain walls in flat space \cite{vis}, which was utilized in the construction of inflating branes in braneworld models \cite{branes}.
The `appearance' of staticity in (\ref{so31}) is therefore just a mirage, coming from the codimension-1 structure of the bubble wall. In the reference frame of an inertial observer at infinity, however, one can easily ascertain that the bubble is expanding.

To find the field profiles in (\ref{so31}), we can again work in the action as before. It is then convenient to change the gauge (i.e. `un-fix' the gauge) in (\ref{so31}) by defining new `radial' coordinate, by
$dz = e^{-\theta} d\zeta/\zeta$, where $\theta$ is an arbitrary function, and redefine the field $\pi$ to
$\tilde \pi = \pi + \ln \zeta$. Transforming the ansatz (\ref{so31}) to this gauge, and substituting it into the  the action (\ref{action}), after factoring out the common powers of $M_{Pl}$, we get
\ba S&=&\int d^3x \sqrt{-\gamma} \Bigl\{ \int dz \Bigl[3 e^{2\tilde \pi} \Bigl(e^{-\theta}  ( \frac{d\tilde \pi}{dz})^2
-2  \frac{d \tilde \pi}{dz} + e^{\theta} \Bigr)-\frac{1}{2}e^{2\tilde \pi-\theta} (\frac{d\varphi}{dz})^2 -
2\alpha e^{-3\theta} (\frac{d\varphi}{dz})^3 \frac{d\tilde \pi}{dz}\Bigr] \nonumber \\
&&- \sigma e^{3(\tilde \pi+\beta\varphi/2)}|_{z=z(\zeta_w)} \Bigr\} \, ,
\label{so31act}
\ea
where we have performed some integration by parts, taking care of  any relevant boundary terms. The second line of (\ref{so31act}) is the bubble wall tension term at $z=z(\zeta_w)$, with physical tension $\hat \sigma = M_{Pl}^2 \sigma$, given our normalizations. To obtain the field equations governing the configuration (\ref{so31}),
we now vary this action with respect to $\varphi$, $\theta$ and $\tilde \pi$, and after obtaining the field equations, gauge-fixing to $\theta = 0$, $\tilde \pi = \pi + \ln \zeta$, eliminating $z$ for $\zeta$, and integrating the galileon $\varphi$ field equation once, we obtain, respectively,
\ba
&& e^{2\pi} \varphi'+6 \alpha \varphi'^2 \left(\frac{1}{\zeta}+ \pi' \right) = \frac{Q}{\zeta^3} \, , \nonumber \\ 
 &&6 \zeta^3 e^{2\pi} \pi'+\zeta^4 \left[3e^{2\pi} \pi'^2-\frac{1}{2} e^{2\pi} \varphi'^2-6  \alpha \varphi'^3  \left(\frac{1}{\zeta}+\pi'\right) \right]= 0 \, , \nonumber \\
&& \frac{1}{\zeta^3} \left( \zeta^3 e^{2\pi} \pi'\right)'=\frac{ \alpha}{3\zeta^3} \left( \zeta^3  \varphi'^3 \right)'+ \frac16 e^{2\pi} \left(6 \pi'^2-\varphi'^2\right) \, ,
 \label{phieom}
\ea
off the domain wall, where, as before, $Q$ is an integration constant which follows from shift 
symmetry $\varphi \to \varphi+\textrm{constant}$. Primes designate $\zeta$-derivatives. On the wall, the 
the junction conditions relating the sections of (\ref{so31}) inside and out are given by the continuity conditions $\Delta \pi=\Delta \varphi=0$, and the analogue of the Israel equations 
\ba
&&\Delta \left( e^{2\pi} \pi'-\frac{ \alpha}{3 } \varphi'^3  \right) = -\frac{ \sigma}{2} e^{3(\pi+\beta\varphi/2)}  \, ,
\nonumber \\
&&\Delta \left(\frac{ Q}{\zeta^3} \right) =  \frac{3 \beta \sigma}{2} e^{3(\pi+\beta\varphi/2)} \, ,
\label{bcso31}
\ea
where as before  $\Delta$ represents the difference of an appropriate quantity between the outside and the inside of the wall. These are easily obtained from (\ref{phieom}) with the wall contributions included as $\delta$-function terms following from varying (\ref{so31act}), and using the Gaussian pillbox integration, as before.
Further, as we discussed in previous sections, on the wall we can always absorb the value of $\varphi$ into the definition of the matter coupling $e^{\beta \varphi}$, and so we can always set $\varphi=0$ on the wall. 

Exact solutions of this system of equations are not known to date. However we can employ an expansion in  $1/M_{Pl}$, and substitute into (\ref{phieom}), (\ref{bcso31}) the perturbative ansatz
\be
\pi =\frac{\pi_0}{M_{Pl}}+\frac{\pi_1}{M_{Pl}^2}+\ldots, \qquad \varphi =\frac{\varphi_0}{M_{Pl}}+\frac{\varphi_1}{M_{Pl}^2}+\ldots,\qquad  \sigma =\frac{\sigma_0}{M_{Pl}}+\frac{\sigma_1}{M_{Pl}^2}+\ldots \, ,
\ee 
and recall that $\alpha=M_{Pl} \hat \alpha$. To leading order we recover the decoupling limit in which $M_{Pl} \to \infty$ with $\hat \alpha$ and $\hat \sigma/M_{Pl}$ held fixed. The higher order gravity corrections are then obtained iteratively, order-by-order in $M_{Pl}^{-1}$. After the calculation is done, we can absorb the one extra power of $M_{Pl}^{-1}$ to render the leading order solutions dimensionless, as per our conventions.
The key is to find the solutions which are smooth on either side of the bubble. The reason is that such solutions have smooth Euclidean continuation, and so one would expect to have an instanton describing their nucleation which has a finite action. This is made easier when one turns on matter vacuum energies inside and outside the bubble, and allows them to dominate asymptotically far from the bubble. In that case, one will find instantons which are merely ``squashed" standard tunneling solutions in General Relativity, looking like creased spheres. In our case, what suffices is to demonstrate that even without matter vacuum energies nonsingular configurations exist. We will not go after determining their action here, although that task warrants further investigation.

The configuration describing a bubble of self-accelerating branch in the vacuum of the asymptotically flat branch in the interior has the fields
\be
\pi^{in}(\zeta)=\frac{1}{48(6 \alpha)^2}(\zeta_w^4-\zeta^4)+{\cal O}(M_{Pl}^{-3}) \, , \qquad \frac{d\varphi^{in}}{d\zeta}=-\frac{\zeta}{6 \alpha}+{\cal O}(M_{Pl}^{-3}) \, ,
\ee
while outside they are
\be
\pi^{out}(\zeta)={\cal O}(M_{Pl}^{-3}), \qquad \frac{d\varphi^{out}}{d\zeta}=-\frac{\beta \zeta_w^6}{12 (6 \alpha)^2 \zeta^3}+{\cal O}(M_{Pl}^{-3}) \, .
\ee
In deriving these solutions, we have imposed regularity at $\zeta=0$, and the asymptotic boundary condition $\pi^{out} \to 0$. This suffices since away from the wall we can always shift $\varphi$ by a constant, as we explained above.
The  galileon charges are given by
\be 
Q^{in}={\cal O}(M_{Pl}^{-3}) \, , \qquad Q^{out}=-\frac{\beta \zeta_w^6}{12  (6 \alpha)^2 }+{\cal O}(M_{Pl}^{-3}) \, ,
\ee
whereas the brane tension is
\be
\sigma=-\frac{\zeta_w^3}{18 (6 \alpha)^2 }+{\cal O}(M_{Pl}^{-3}) \, .
\ee
For any finite size bubble, the  tension is {\it negative}. This suggests that to nucleate the bubbles of the self-accelerating branch in the asymptotically flat vacuum of the normal branch may be difficult, suggesting the need for {\it ghost} matter fields to account for the negative tension. Specifically, if we were to try to associate the tension only to the wall-localized matter, we'd need to put negative energy modes there. On the other hand, since there is a ghost already in the theory, on the self-accelerating branch, one might imagine that in some resolved, thick wall description of the transition, the kink of such a ghost induced by the bubble wall of normal matter, might provide the requisite negative energy. Indeed, the scaling of the field gradients in the solution with the size of the bubble are curiously in accord with such a possibility. However, without a more detailed demonstration that such kinks may actually arise, which is beyond the scope of the present work, it is impossible to pass the immediate verdict on the transitions in this direction. Furthermore, when the matter vacuum energies are turned on, this problem will get even more complicated thanks to the difference of the vacuum energy on different sides of the wall to the wall's tension. Those contributions may even make the overall tension positive, and allow for transitions to the self-accelerating branch even without negative energies on the wall, or near it. This issue deserves further scrutiny, and we hope to return to it in the future. 

Describing the reverse transitions is simpler. The configuration describing bubbles of the flat vacuum expanding in the self-accelerating background are given by the field solutions 
in the interior,
\be
\pi^{in}(\zeta)={\cal O}(M_{Pl}^{-3}) \, , \qquad \frac{d\varphi^{in}}{d\zeta}={\cal O}(M_{Pl}^{-3}) \, ,
\ee
which are surrounded by a self-accelerating exterior\footnote{To see explicitly why this is the self-accelerating solution, note that the physical metric, as seen by matter,  is $\hat g_{\mu\nu}=e^{\beta \varphi} g_{\mu\nu}$ which to leading order in our expansion goes like $\hat g_{\mu\nu} \simeq\left( 1-\frac{\beta \zeta^2}{12\alpha}\right) \eta_{\mu\nu}$ in the exterior region, where $\zeta^2=\eta_{\mu\nu}x^\mu x^\nu$.  For $\alpha\beta>0$, this is just de Sitter space written as a local perturbation about Minkowski, along the lines discussed in section 1.1 of \cite{galileon}.}
\be
\pi^{out}(\zeta)=\frac{1}{48(6 \alpha)^2}(\zeta_w^4-\zeta^4)+{\cal O}(M_{Pl}^{-3}), \qquad \frac{d\varphi^{out}}{d\zeta}=-\frac{\zeta}{6 \alpha}-\frac{\beta \zeta_w^6}{12 (6  \alpha)^2 \zeta^3}+{\cal O}(M_{Pl}^{-3}) \, .
\ee
Again, we have imposed regularity at $\zeta=0$, and picked $\pi(0)=0$ by scaling $\zeta$. The  galileon charges are now  given by
\be 
Q^{in}={\cal O}(M_{Pl}^{-3}), \qquad Q^{out}=\frac{\beta \zeta_w^6}{12 (6 \alpha)^2 }+{\cal O}(M_{Pl}^{-3}) \, ,
\ee
and  the brane tension is
\be
\sigma=\frac{\zeta_w^3}{18 (6 \alpha)^2 }+{\cal O}(M_{Pl}^{-3})
\ee
This time the brane tension is {\it positive} for $\zeta_w>0$.  Therefore transitions from the self-accelerating branch to the normal branch do not require any exotic types of matter to catalyze them. They are mediated by the coupling between the {\it unstable} gravitational sector and the ordinary matter fields whose excitations ultimately populate the bubble wall. We expect that this is a non-perturbative manifestation of the perturbative ghost instability on the self-accelerating branch. Note, of course, that at zeroth order in our expansion, we recover the result of \cite{koyama-tanaka-pujolas} in the decoupling limit, where the tension is found to be zero. Here we see that the  leading order gravity correction to that is one of {\it positive} tension. 

To sum up this section, we see that there {\it exist} channels for nonperturbative branch changing. A branch change can be accomplished locally and in a causal manner, by a nucleation of a nonsingular ``hairy" bubble,
whose wall kicks the galileon field and while sourcing its hair, it pushes it do a different branch from the parent configuration. Bubble walls which demarcate self-accelerating vacua in the interior from the external asymptotically flat vacuum require a negative tension, and further analysis is needed to determine if they can naturally arise in the theory. The bubbles which describe sections of an asymptotically flat vacuum in the interior from the external self-accelerating vacuum however have walls with positive tension, and so may be constructed using the standard matter sources. It also remains to see what happens when we allow for nonzero matter contributions to the vacuum energy inside and out of the bubble, and whether these branch changing bubbles really provide the means for exploring the full structure of the landscape of modified gravity, a view argued for in \cite{kalkil}.

\section{Summary}

In this work, we have been able to give for the first time the complete proof of the Vainshtein mechanism, which is widely assumed to operate in modified gravity models with strongly coupled additional modes in the gravitational spectrum. Working with a simple cubic galileon theory and thin shell spherically symmetric static sources, a.k.a. Dyson spheres, we find the first integrals to all the field equations, reducing them to a more tractable first order system. This allows one to classify all the solutions, and specifically show that for a range of matter sources on the Dyson spheres, there exist asymptotically flat solutions with scalar hair. The scalar hair
scaling with distance changes from the standard $4D$ form,  $\phi \sim 1/r$
dictated by the usual $4D$ Gauss law, to another scaling, controlled by the higher-dimension operator in the galileon sector, at a very  large distance, the Vainshtein radius. This is the key to how the Vainshtein mechanism manages to hide the extra forces induced by the additional light modes in the theory from probe particles moving inside the Vainshtein
radius. The hair changes by getting fuzzy below the Vainshtein radius, and so its field lines do not bunch up as densely as in the usual $4D$ case, remaining more dilute. Because this transition occurs far, where the fields are weak, they remain weak all the way very close to the source. Since the force acting on a probe is still $\propto \phi'$, it is therefore also very weak. 

For $\alpha\beta>0$,
 this Vainshtein coiffure, of fuzzy hair below the Vainshtein radius, sets up for matter sources on the shell whose energy density and pressure obey $\rho > 2p > -
\rho$ and $\rho > 0$. This includes much of the usual matter, such as radiation and non-relativistic mass distributions. However, when pressure is large, $p > \rho/2$, as would be the case if the matter is field gradients confined to the shell, the hair would not splinter efficiently, and the Vainshtein effect would not be operable. 
Similarly, for $\alpha\beta<0$, the Vainshtein coiffure is only styled for matter sources on the shell  satisfying $0<\rho <2p$, and $\rho+2p<0$. It is immediately evident that this rules out shells composed of ordinary pressureless matter, or even a combination of matter and radiation whose effective equation of state lies somewhere in between $w=0$ and $w=1/2$. 
All of this points to the general issue of sensitivity of the Vainshtein effect on the details of the source distribution, and we indeed find signs of that. Specifically, we show that the Vainshtein radius depends not only on the total mass of the source, but also its properties, as encoded by the pressure. 

We expect that the above considerations can be generalized beyond our Dyson spheres to more complicated matter configurations. The condition $\rho+2p > 0$ is just   the strong energy condition,  while $\alpha \beta(\rho-2p) >0$ is equivalent to requiring the trace of energy momentum to take a particular sign (depending on the sign of $\alpha\beta$), $\alpha\beta T<0$.  The latter condition is particularly important as it controls whether or not the Vainshtein mechanism is realized ($\alpha \beta T<0$) or if the solution is expected to run into a singularity ($\alpha \beta T>0$). Note that in the marginal case of $T=0$ the galileon is not sourced at any scale, and it is subtle to decide if the Vainshtein effect is present at all. On the one hand, the absence of any galileon effects may suggest that for this case the Vainshtein mechanism is extremely efficient. On the other hand, as we see from Eq. (\ref{preciserv}) the Vainshtein radius vanishes for such sources, implying that the Vainshtein effect is maximally inefficient. The point is, that for such sources there is nothing to suppress in the first place, since they are bald to start with - the absence of any hair makes their fields automatically the same as in conventional General Relativity. Thus, in reality, for this special limit, the Vainshtein effect is irrelevant.

In addition, we find that generically the theory admits more solutions, which in the static limit are all singular far away from the Dyson sphere. In many cases, one expects that those solutions cannot be regulated, and so may have to be thrown out of the theory, in a way similar to what happens to their relatives in conventional General Relativity\footnote{It should be noted that the issue of singular solutions in massive gravity has been recently explored also in \cite{nyu,gruzinov2}. We believe that those issues are similar to what one encounters in galileon systems.}.
However, one of the singular solutions can be readily regulated by turning on time dependence, since it is really the self-accelerated, ghost infested, branch familiar in galileon models. We explore what happens if we try to combine such time dependent solutions with shells laden with sources whose pressure and energy density obey $\rho + 2p <0$, which corresponds to the scalings of shell-bound `dark energies'. In this case, we find 
regular solutions, where the shells expand separating regions of different branch vacua. These configurations  point that the self-accelerating branch may decay to the asymptotically flat branch. The case where the bubbles describe pockets of self-accelerating branch inside the asymptotically  Minkowski vacuum are more complicated because they require negative tension, and further work is needed to determine what really transpires.  Generally, our constructions point to channels for nonperturbative transitions between branches, which need to be addressed further to determine phenomenological viability of multi-branch gravities.

\vskip.5cm

{\bf \noindent Acknowledgements}
 
\smallskip

We thank Mina Arvanitaki, Savas Dimopoulos, Takahiro Tanaka and Giovanni Villadoro for useful discussions.
AP thanks the UC Davis HEFTI program
for hospitality during the inception of this work. The research of NK and NT is
supported in part by the DOE Grant DE-FG03-91ER40674.  AP is funded by a Royal Society University Research Fellowship.



\end{document}